\documentclass[aps,prx,a4paper,nofootinbib,superscriptaddress,numbers,longbibliography,showpacs,showkeys,floatfix,svgnames,reprint]{revtex4-2}

\usepackage{amsfonts,amssymb,amsmath}
\usepackage{hyperref,xcolor,graphicx}
\usepackage[separate-uncertainty=true]{siunitx}

\DeclareSIUnit\gauss{G}

\usepackage[english]{babel}
\usepackage[utf8]{inputenc}
\usepackage[T1]{fontenc}

\usepackage{xspace}

\usepackage{physics}

\newcommand{\figref}[2]{\hyperref[#1]{\ref{#1}{(#2)}}}
\addto\captionsenglish{}

\def\theTitle{%
Direct measurement of the Wigner function of atoms in an optical trap
}

\hypersetup{
 pdftitle = {\theTitle},
 pdfauthor = {Falk-Richard G. Winkelmann, Carrie A. Weidner, Gautam Ramola, Wolfgang Alt, Dieter Meschede, Andrea Alberti},
	colorlinks=true,linkcolor=blue,citecolor=blue,filecolor=blue,urlcolor=blue,pdfpagemode=UseNone,pdfstartview={XYZ null null 1.00}
}

\renewcommand\textemdash{\leavevmode\unskip\kern0.8pt---\kern1pt\ignorespaces}

\makeatletter
\let\old@Section@Cmd=\section
\def\end@of@sec@tit{.\leavevmode\unskip\kern0.8pt\rule[0.19\baselineskip]{8pt}{0.4pt}\kern1pt}
\def\mysection{%
	\def\reserved@stsec##1{\@startsection{section}{1}{\parindent}{\z@}{0em}{\normalfont\normalsize\itshape}*[##1]{##1\end@of@sec@tit}}%
	\@ifstar{%
		\reserved@stsec%
	}{%
		\reserved@stsec%
	}%
}%
\let\section=\mysection
\makeatother

\begin{document}
\title{\theTitle}

\author{Falk-Richard Winkelmann}
\affiliation{Institut für Angewandte Physik, Universität Bonn, Wegelerstraße 8, D-53115 Bonn, Germany}
 \author{Carrie A. Weidner}
\affiliation{Quantum Engineering Technology Laboratories, H. H. Wills Physics Laboratory and Department of Electrical and Electronic Engineering, University of Bristol, Bristol BS8 1FD, United Kingdom}
\author{Gautam Ramola}
\affiliation{Institut für Angewandte Physik, Universität Bonn, Wegelerstraße 8, D-53115 Bonn, Germany}
\author{Wolfgang Alt}
\affiliation{Institut für Angewandte Physik, Universität Bonn, Wegelerstraße 8, D-53115 Bonn, Germany}
\author{Dieter Meschede}
\affiliation{Institut für Angewandte Physik, Universität Bonn, Wegelerstraße 8, D-53115 Bonn, Germany}
\author{Andrea Alberti}
 \email{alberti@iap.uni-bonn.de}
\affiliation{Institut für Angewandte Physik, Universität Bonn, Wegelerstraße 8, D-53115 Bonn, Germany}

\date{\today}             
\begin{abstract}
	We present a scheme that uses Ramsey interferometry to directly probe the Wigner function of a neutral atom confined in an optical trap.
	The proposed scheme relies on the well-established fact that the Wigner function at a given point $(x,p)$ in phase space is proportional to the expectation value of the parity operator relative to that point.
	In this work, we show that parity-even and parity-odd motional states can be mapped to two distinct internal states of the atom by using state-dependent trapping potentials.
	The Wigner function can thus be measured point-by-point in phase space with a single, direct measurement of the internal state population.
	Numerical simulations show that the scheme is robust in that it applies not only to deep, harmonic potentials but also to shallower, anharmonic traps.
\end{abstract}

                              \maketitle

\section{Introduction}
\label{sec:introduction}

A basic principle of quantum mechanics is that the motional state of a quantum particle is completely described by a density matrix with complex entries.
It was shown by Wigner \cite{Wigner1932} that the same information contained in the density matrix can be encapsulated in a real-valued function in phase space.
The Wigner function thus establishes a complete representation of the motion of a quantum particle in phase space \cite{Groenewold1946,Moyal1949}.

A unique property of this phase-space representation is that the marginal distributions of the Wigner function reduce to the probability densities in real and momentum space \cite{Hillery1984,Zachos2005}.
In contrast to classical probability distributions, however, the Wigner function of quantum states can take negative values.
This is the case for all pure, non-Gaussian states \cite{Hudson_1974}.
One can indeed interpret negative values of the Wigner function as an indicator of the state's non-classicality \cite{Kenfack_2004}.
Furthermore, unlike the density matrix, the Wigner representation conveys visual and insightful information about the motion of a quantum particle.
This visual character of the Wigner function, combined with the fact that it applies to both the classical and quantum worlds, lends it ideally to the study of decoherence phenomena \cite{Deleglise2008}.
Due to these properties, the Wigner function is a useful tool in many research areas ranging from fundamental studies in quantum optics to applications in quantum information science and quantum sensing~\cite{Weinbub2018}.

\begin{figure*}[t!]	
	\centering 
	\includegraphics[width=\textwidth]{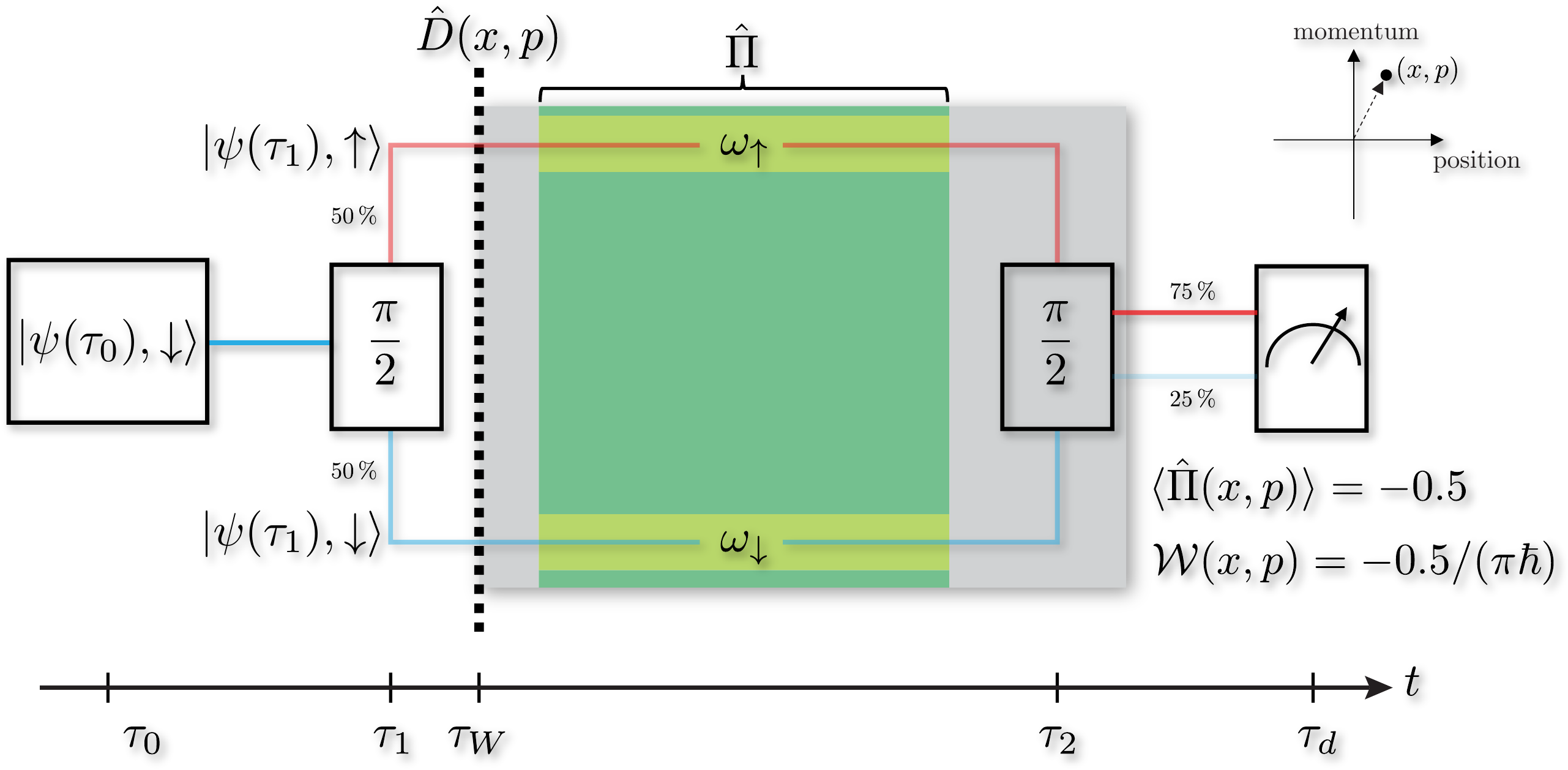}
	\caption{Schematic illustration of the Ramsey interferometric scheme. At time $\tau_0$, a motional quantum state $\ket{\psi(\tau_0)}$ is prepared in a known spin state $\ket{\downarrow}$. Subsequently a $\pi/2$ pulse splits the state into an equal superposition at time $\tau_1$, 
	with each component of the superposition represented by blue and red lines, respectively, with the transparency reflecting their relative population.
	At the time when the Wigner function is to be measured, $\tau_W$, the potential is shifted abruptly (thick, black, dashed line) by a distance $x$ and subsequently kept moving (grey shaded region) with constant speed $p/m$, with $m$ being the mass of the atom, until a second $\pi/2$ pulse interferes the two states at time $\tau_2$.
		Between the two $\pi/2$ pulses, the spin-dependent potential is changed for some for some probing time (shaded green-yellow regions) such that the atom experiences a different trap frequency, $\omega_\uparrow$ and $\omega_\downarrow$, depending on its internal state.
	The differential trap frequency and probing time are chosen such that the relative phase accrued by a given Fock state $\ket{n}$ is $n\pi$, thus projecting states of opposite parity onto opposite poles of the Bloch sphere when the internal state is finally measured at time $\tau_d$.
	 	In the example illustrated here, a negative Wigner function $\mathcal{W}(x,p)=-0.5/(\pi\hbar)$ is measured.}
	\label{fig:Wigner_Idea}
\end{figure*}

There are two main approaches to obtain the Wigner function experimentally.
These are denoted as direct or indirect, depending respectively on whether or not the Wigner function $\mathcal{W}(x,p)$ can directly be obtained at a specific point $(x,p)$ in phase space \cite{Leibfried1997}.

In an indirect approach,
one first reconstructs the entire quantum state, e.g., by determining the density matrix, and subsequently derives $\mathcal{W}(x,p)$ through a mathematical transformation.
A plethora of techniques have been developed to first determine the quantum state by measuring the Husimi Q function \cite{Lv:2017}, probing other phase-space representations \cite{Deleglise2008}, tomographically recording rotated quadratures \cite{Poyatos1966,Lvovsky:2001}, or determining the characteristic function \cite{Fluhmann:2020}.

When using a direct approach, in contrast, one probes the Wigner function point-by-point in phase space.
For this purpose, it is important that the Wigner function $\mathcal{W}(x,p)$ can be defined operatively in terms of an observable \cite{Cahill1969a,Cahill1969}, which is given by the parity operator $\hat{\Pi}$ displaced to the phase-space point $(x,p)$ \cite{Grossmann1976,Royer1977}: 
\begin{equation}
\label{eq:wigner_def_parity_op}
\mathcal{W}(x,p)=\frac{1}{\pi\hbar}\braket{\psi}{\left.\hat{D}(x,p)\,\hat{\Pi}\, \hat{D}(x,p)^{\dagger}\right|\psi}.
\end{equation}
Here, $\ket{\psi}$ is the quantum state,
and $\hat{D}(x,p) = \exp[i(p \, \hat{x} - x \, \hat{p})/\hbar]$
is the Glauber displacement operator, shifting the position and momentum by $x$ and $p$, respectively.
The operators 
$\hat{x}$ and $\hat{p}$ are the canonical position and the momentum operators.
The value of $\mathcal{W}(x,p)$ is directly obtained if one can measure the expectation value of the displaced parity operator, $\hat{\Pi}_{x,p} = \hat{D}(x,p)\,\hat{\Pi}\, \hat{D}(x,p)^{\dagger}$.

For a trap potential that is invariant under $\hat{\Pi}$, the displaced parity operator $\hat{\Pi}_{x,p}$ takes a simple, diagonal form in the basis of the displaced Fock states $\ket{n}$:
\begin{equation}
	\label{eq:parity_operator}
    \hat{\Pi}_{x,p}= 
    \sum_{n=0}^{\infty} e^{i\pi n}   \hat{D}(x,p) |n\rangle \langle n|  \hat{D}(x,p)^\dagger.
\end{equation}
The effect of this operator on the displaced vibrational states can thus be understood as a phase shift of $\pi$ for the odd states and zero for the even states.
A review of the operator $\hat{\Pi}$ as a quantum mechanical observable is provided in Ref.~\cite{Birrittella2021}.

Based on Eqs.~(\ref{eq:wigner_def_parity_op}) and (\ref{eq:parity_operator}), a direct measurement of the Wigner function can thus be accomplished by first measuring the state occupation probabilities $Q_n(x,p)  = |\langle{\psi}|\hat{D}(x,p)  \ket{n}\hspace{-1pt}|^2$ of a set of displaced vibrational states, and post-processing the measured data to evaluate the sum $\mathcal{W}(x,p) = \sum_{n=0}^\infty (-1)^{n}Q_n(x,p)/(\pi\hbar)$.
This method was experimentally demonstrated, e.g., with trapped ions \cite{Leibfried:1996} and coherent states of light \cite{Banaszek1999}.
Alternatively, one can obtain the expectation value of $\hat{\Pi}_{x,p}$ in Eq.~(\ref{eq:parity_operator}) in one single measurement by using Ramsey interferometry. This requires no post-processing of the measured data, as originally proposed by Lutterbach and Davidovich~\cite{Lutterbach1997} for reconstructing the Wigner function of trapped ions and the electromagnetic field of a microwave resonator.
The second of these two schemes was demonstrated experimentally, reconstructing the Wigner function of the first two Fock states of microwave photons \cite{Nogues2000,Bertet2002}.
Nonlinear interactions between the electromagnetic field and the Rydberg atoms that are used to measure the field parity, however, limit the measurement fidelity when the same scheme is applied to higher-order Fock states \cite{Deleglise2008}.
Subsequent experiments with microwave photons coupled to a superconducting transmon qubit showed that nonlinear distortions can be greatly suppressed in a strongly dispersive regime \cite{Vlastakis:2013}.
For neutral atoms, indirect schemes based on the tomographic reconstruction of the Wigner function have been demonstrated with atoms in free space \cite{Kurtsiefer1997} and, very recently, in optical tweezers \cite{Brown2022}. However, to date, no counterpart scheme has been put forward for the direct reconstruction of the Wigner function.

In this paper, we propose a scheme to directly measure the Wigner function of single neutral atoms trapped in an optical potential.
Our scheme employs a Ramsey interferometer to measure the expectation value of the parity operator, closely resembling the seminal method developed by Lutterbach and Davidovich \cite{Lutterbach1997}.
The basic idea of the proposed scheme is illustrated in Fig.~\ref{fig:Wigner_Idea}.
We denote the atom's motional state with $\ket{\psi}$ and its internal state as either $\ket{\uparrow}$ and $\ket{\downarrow}$; that is, the atom's internal state is a two-level system. We start with the atom at time $\tau_0$ in a given motional and internal state, $\ket{\psi(\tau_0)}\ket{\downarrow}$, and then create at time $\tau_1$ a quantum superposition of two copies of the atom's motional state, each associated with a different internal state.
At a later time $\tau_W$, a sudden displacement of the trap position jointly with an abrupt change of its velocity realizes the displacement operator, producing the state $D^\dagger(x,p)\ket{\psi(\tau_W)}(\ket{\downarrow} + \ket{\uparrow})/\sqrt{2}$ in the reference frame comoving with the trap.
The atom is then subject to a state-dependent trap potential in which the two original copies follow different evolutions.
By carefully modulating the state-dependent potential, one copy becomes the mirror image (i.e., the parity symmetry partner) of the other, thus creating effectively 
an equal superposition of $D^\dagger(x,p)\ket{\psi(\tau_W)}\ket{\downarrow}$ and $\hat{\Pi}\,D^\dagger(x,p)\ket{\psi(\tau_W)}\ket{\uparrow}$.
Finally, by letting these two states interfere at time $\tau_2$, we obtain the expectation value of the parity operator from the (signed) contrast of the Ramsey fringe, \raisebox{0pt}[0pt][0pt]{$C(x,p)=\braket*{\psi(\tau_W)}{\hat{\Pi}_{x,p}|\psi(\tau_W)}$}, which can be measured without scanning the whole fringe, as will be shown later.
Hence, the Wigner function measured at the phase-space point $(x,p)$ can be simply expressed as $\mathcal{W}(x,p) =C(x,p)/(\pi\hbar)$.

In what follows, we elucidate the scheme in more detail by first considering the ideal case of a harmonic oscillator.
We then proceed by discussing the more realistic scenario of an anharmonic potential, focusing on neutral atoms trapped in a deep optical lattice potential.
We show that the effect of the trap anharmonicity is greatly suppressed because the nonlinearity of the optical potential affects, to first-order, both spin states equally, and thus yields no contribution to the Ramsey interferometric signal.
We will discuss how these results, including the common-mode suppression of the anharmonicity, can be adapted to neutral atoms trapped in optical tweezers.\\

\section{Parity measurement in an ideal harmonic potential}

We consider the ideal situation of a harmonic potential, where the eigenenergies \raisebox{0pt}[0pt][0pt]{$E^\mathrm{(HO)}(n)$} depend linearly on the harmonic trap frequency $\omega$:
\begin{equation}
    E^\mathrm{(HO)}(n) = \hbar \omega \left(n+\frac{1}{2}\right)-U_0,
\end{equation}
Here, the term $U_0$ accounts for the trap depth, i.e., the energy shift caused by the trapping potential at its center.
This shift occurs in the typical situation of an optical dipole trap that is red detuned with respect to the atomic resonance.

\begin{figure}[t]	
	\centering
    \includegraphics[width=\columnwidth]{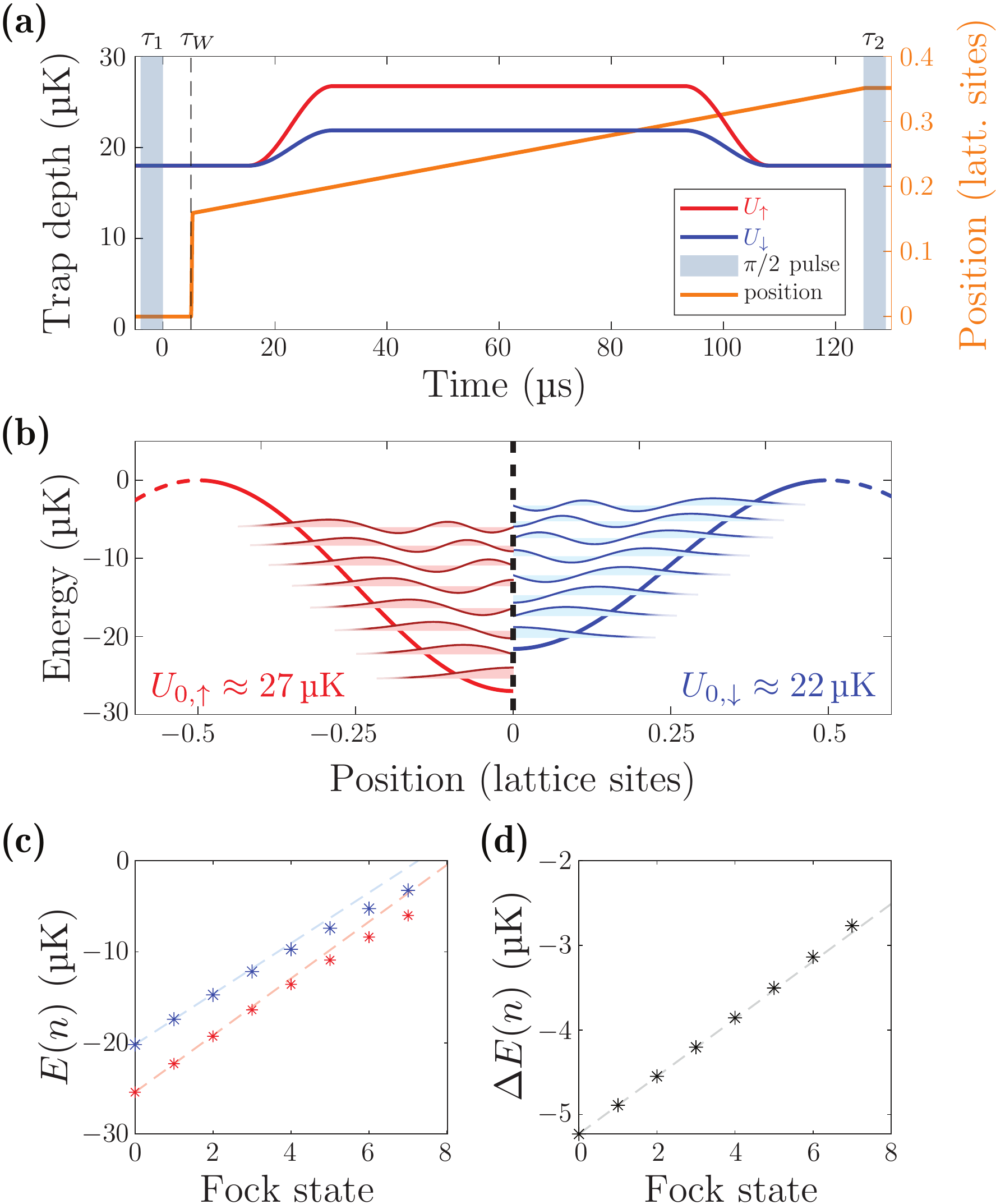}
	\caption{(a) Schematic illustration of the Ramsey interferometric sequence for measuring the Wigner function, showing the two $\pi/2$ Ramsey pulses, the trap depths (left axis) for the two internal states, $U_{\uparrow}$ and $U_\downarrow$, and the trap displacement (right axis), which begins at $\tau_W$.
	For reference, the trap period is $2\pi/\omega \approx \SI{18}{\micro\second}$ at time $t=0$.
	(b) The wavefunctions of the motional Fock states, vertically shifted by their motional eigenenergies, are shown for the two internal states, together with the corresponding lattice potentials, on the left- and right-hand side, respectively.
	Owing to parity symmetry, the illustration would extend symmetrically for each internal state into the other half-plane.
	Note that only states with negligible tunneling probability are shown, and that the internal state energy is not considered in the illustration.
	(c) A plot of the eigenenergies for the two internal states, $E_\uparrow(n,t)$ (red points) and $E_\downarrow(n,t)$ (blue points), with lattice depths $U_{0,\uparrow} \approx k_B\times\SI{27}{\micro\kelvin}$ and $U_{0,\downarrow} \approx k_B\times\SI{22}{\micro\kelvin}$, respectively ($k_B$ is the Boltzmann constant).
	The spacing between adjacent levels is not constant due to the potential anharmonicity.
	For reference, the spectra of a harmonic oscillator are shown with dashed lines.
	(d) The differential light shift $\Delta E_n$ is plotted vs.\ the Fock state $n$, with the dashed line showing the case of an ideal harmonic oscillator.
	Deviations, expressed in relative units of $\hbar\omega$, are below $0.1$ for the lowest six vibrational levels, owing to the first-order common-mode suppression of the anharmonic contribution.
	}
	\label{fig:Wigner_EigenenergiesUnbalancedLattice}
\end{figure}

Figure~\figref{fig:Wigner_EigenenergiesUnbalancedLattice}{a} provides a schematic illustration of the operations for reconstructing the Wigner functions.
To begin with, we describe the interferometric Ramsey sequence, which allows us to measure the expectation value of the parity operator $\hat{\Pi}$.
As shown in Fig.~\ref{fig:Wigner_Idea} and described above, with a first $\pi/2$ pulse, the atom is prepared in an equal superposition of the two internal states, $\ket{\uparrow}$ and $\ket{\downarrow}$.
We then use state-dependent potentials \cite{Deutsch:1998,Robens2018} to independently vary the trap depths, $U_\uparrow$ and $U_\downarrow$, that the atom experiences depending on its internal state.
In the case of an adiabatic modulation of the trap depth, for any given Fock state $\ket{n}$, the accumulated phase difference between the two internal states during the Ramsey sequence is given by the time integral
\begin{equation}
\label{eq:diff_phase}
\Phi(n) = \int_{\tau_1}^{\tau_2}\hspace{-3pt}{\rm {d}t}\, \Delta E^\mathrm{(HO)}(n,t)/\hbar.
\end{equation}
Here, $ \Delta E^\mathrm{(HO)}(n,t)$ represents the differential energy, defined as the difference between the instantaneous eigenenergies, \raisebox{0pt}[0pt][0pt]{$E_\uparrow^\text{(HO)}(n,t)$} and \raisebox{0pt}[0pt][0pt]{$E_\downarrow^\text{(HO)}(n,t)$}, for the two internal states.
The differential energy consists of two terms,
\begin{equation}
    \Delta E^\textrm{(HO)}(n,t) = \hbar \Delta\omega(t)\, n + \Delta E_0(t),
    \label{eq:DeltaE_HO}
\end{equation}
where one term is proportional to $n$ and the other not.
Moreover, $\Delta \omega(t) = \omega_\uparrow(t)-\omega_\downarrow(t)$ denotes the differential trap frequency, and $\Delta E_0(t) = -U_\uparrow(t) + U_\downarrow(t) + \hbar\Delta \omega(t)/2$ represents a differential energy offset, which only depends on the trap depth and not the motional state.
The second term in Eq.~\eqref{eq:DeltaE_HO} contributes to the Ramsey phase $\Phi(n)$ with a fixed phase shift $\Phi_0 = \intop_{\tau_1}^{\tau_2}{\rm {d}t}\, \Delta E_0(t)/\hbar$, independent of $n$.
This spurious contribution can be conveniently cancelled by choosing the phase difference between the first and the second Ramsey pulses to be equal to $-\Phi_0$. 
When this is the case, the accumulated Ramsey phase only depends on the contribution from the first term, which is proportional to $n$.
The aim is to modulate the differential trap frequency $\Delta \omega$ in time so as to satisfy the condition
\begin{equation}
     \Phi(n)  = \int_{\tau_1}^{\tau_2}\hspace{-4pt}{\rm {d}t}\,\Delta\omega(t) = n\pi.
    \label{eq:Wigner_N_PI}
\end{equation}
When this is the case, the second Ramsey $\pi/2$ pulse transfers Fock states of opposite parity to opposing poles of the Bloch sphere.
Hence, a single measurement of the difference $w$ of the internal-state relative populations yields the signed contrast $C = w$ of the Ramsey fringe, from which one readily obtains the Wigner function probed at the origin, $W(0,0) = w/(\pi\hbar)$.

The procedure described so far allows us to measure the parity of the wavefunction and, thus, to probe the Wigner function at the origin.
To probe a generic point $(x,p)$ in phase space, we propose to abruptly change the position and velocity of the optical trap at the instant $\tau_W$ at which we want to record a snapshot of the Wigner function $\mathcal{W}(x,p)$. 
At this point, the trap position is shifted by $x$, and subsequently kept moving at constant speed $p/m$, where $m$ is the mass of the atoms.
By the equivalence principle, in the reference frame co-moving with the optical trap, the change of position and velocity corresponds to a displacement of the quantum state as is implemented by the operator $\mathcal{D}(x,p)^\dagger$ appearing in Eq.~(\ref{eq:wigner_def_parity_op}).
Because the displacement operation is spin-independent, one is free to apply the sudden shift of position and velocity before the first $\pi/2$ pulse, thus inverting the order of $\tau_1$ and $\tau_W$ in Fig.~\ref{fig:Wigner_Idea} without affecting the validity of the scheme.

Three remarks are in order. The first one regards the temporal resolution of the Wigner function measurement.
This is determined by how sharply in time one can shift the position and velocity of the trap.
In fact, to avoid blurring the reconstructed Wigner function, the displacement operation must be faster than the typical timescale of motion in the trap, which is set by the trap oscillation period.
In contrast, the temporal resolution does not depend on how fast the trap frequency is varied during the Ramsey sequence.
Rather, the trap frequencies $\omega_\uparrow$ and $\omega_\downarrow$ must be changed sufficiently slowly, i.e., over a timescale comparable or longer than the harmonic oscillation period, in order to avoid coupling between Fock states.

The second remark pertains to the direct character of the measurement.
The experimental sequence need be calibrated only once, e.g., when probing the origin $(x,p) = (0,0)$ and need not be recalibrated for other points in phase space.
In fact, neither the condition on the relative phase of the two Ramsey pulses, cancelling the contribution $\Phi_0$, nor the condition in Eq.~(\ref{eq:Wigner_N_PI}), realizing the parity operator, depend on the displacement operator.
As a consequence, the physical information about the Wigner function is encoded entirely in the (signed) contrast and not in the phase of the Ramsey fringe, meaning that one can measure the relative population $w$ directly, without scanning the whole fringe.
Moreover, if the measurement is performed on an ensemble of atoms trapped in equal traps, in principle, even a single repetition of the experimental sequence is sufficient to directly determine for any given point $(x,p)$ the relative population $w$ and, thus, $\mathcal{W}(x,p)$.

The third remark deals with the adiabaticity requirement of the parity operation, which differentially modulates the trap frequencies $\omega_\uparrow(t)$ and $\omega_\downarrow(t)$.
In general, for a symmetric trap potential, the parity of the state is preserved regardless of how fast the trap frequencies are varied.
However, parity preservation is not sufficient to exclude parity measurement errors.
One must prevent that for the two internal states, differential motional excitations are created at the end of the Ramsey sequence.
Otherwise, the parity operation applied to a parity-symmetric state (i.e., an eigenstate of the parity operator) produces Ramsey fringes with contrast less than unity. This arises because the two interfering wavefunctions associated with the two internal states consist of different Fock state populations due to these motional excitations, with resulting parity values no longer restricted to the discrete values of $-1$ and $1$.
Such a detrimental effect can be avoided if the trap frequencies are adiabatically varied so to entirely prevent the instantaneous motional eigenstates from coupling to each other;
this is the approach followed in this work.
Alternatively, one could use optimal control methods to determine faster modulation ramps \cite{Weidner:2018}, which, without relying on the adiabatic criterion, can ensure a Ramsey fringe with unit contrast for parity-symmetric states.

\section{Parity measurement in an anharmonic potential}

In any real trap, the energy spectrum exhibits nonlinearity, i.e., the spacing between the levels is not constant.
The reason for this is simply that the depth of any real trap must be finite and therefore its potential becomes anharmonic as the energy increases.
In contrast, an ideal harmonic potential extends indefinitely in space and energy.
In this section, we analyze the effect of such nonlinearity on the measured Wigner function, focusing
specifically on the case of atoms trapped in a single site of a deep optical lattice.
We note, however, that this technique can be generalized to, e.g., atoms in tight optical tweezer traps, as we discuss later.

We consider a one-dimensional optical lattice produced by two counter-propagating laser beams of wavelength $\lambda$.
Its periodic potential can be modeled as
\begin{equation}
\label{eq:potential_latt}
U(x)=-U_0\cos^2\left(\frac{2\pi}{\lambda} x \right),
\end{equation}
where the trap depth $U_0$ is proportional to the intensity of the trapping light and is a positive quantity in the case of a red-detuned dipole light-atom interaction.
From the extended eigenstates (i.e., Bloch states) of the lattice Hamiltonian, one can derive states that are localized to a given lattice site; see Fig.~\figref{fig:Wigner_EigenenergiesUnbalancedLattice}{b}.
These are well-known as Wannier states \cite{Modugno:2012}.

In general, the Wannier states are not eigenstates of the Hamiltonian.
However, they can be considered as such when site-to-site tunneling can be neglected on the timescale of the experiment.
This is the case for low-lying Wannier states in a deep optical lattice, $U_0\gg E_\mathrm{rec}$, where $E_\mathrm{rec} = \left(\hbar k_\lambda\right)^2\slash (2m)$ is the recoil energy for an atom with mass $m$ and a lattice with wavenumber $k_\lambda = 2\pi/\lambda$.

For the symmetric potential in Eq.~(\ref{eq:potential_latt}), the Wannier states have the same parity as Fock states in a harmonic oscillator.
Based on this analogy, we refer henceforth to them as Fock states and use the same notation $\ket{n}$ to denote them.
Figure~\figref{fig:Wigner_EigenenergiesUnbalancedLattice}{c} compares the instantaneous eigenenergies $E_\uparrow(n,t)$ and $E_\downarrow(n,t)$ for the two internal states, relative to the harmonic case, at the point where the differential trap frequency $\Delta \omega(t)$ has reached its maximum.
The plotted eigenenergies are obtained by numerical diagonalization of the lattice Hamiltonian where the momentum and position operators are discretized on a sufficiently fine grid, and the lattice is modelled as a single site with periodic boundary conditions.
The comparison shows deviations from a linear spectrum that increase for larger $n$.
These deviations can be estimated by computing the energy spectrum using first-order perturbation theory \cite{Blatt:2009},
\begin{equation}
\label{eq:EnergySpacingLattice}
E_n = E^\mathrm{(HO)}_n-\frac{2n(n+1)+1}{4}E_\text{rec} + \hbar\omega\,\mathcal{O}\left(E_\text{rec}^2/(\hbar\omega)^2\right),
\end{equation}
where $\omega = k_\lambda\sqrt{2U_0/m}$ represents the trap frequency.
While the term that is quadratic in $n$ in Eq.~(\ref{eq:EnergySpacingLattice}) becomes rapidly significant as $n$ increases, this term has a much smaller effect on the differential energy $\Delta E(n)$.
The reason is that the quadratic term, being proportional to $E_\text{rec}$, does not depend on the trap depth and, thus produces the same contribution for both internal states.
As a result, we expect that, to first order in perturbation theory, $\Delta E(n,t) = E_\uparrow(n,t) - E_\downarrow(n,t) \approx \Delta E^\text{HO}$.
This common-mode suppression of nonlinear distortions is confirmed by a numerical calculation of the differential energy, as plotted as a function of $n$ in Fig.~\figref{fig:Wigner_EigenenergiesUnbalancedLattice}{d}.
The figure shows that deviations from the harmonic case are much smaller than one might expect. 

\section{Experimental setup}
In what follows, for concreteness, we consider the experimental system described in Ref.~\cite{Lam2021a}, where
$^{133}$Cs atoms are trapped in a one-dimensional lattice potential generated by two counter-propagating laser beams with wavelength $\lambda =$ \SI{866}{\nano\meter}.
The two internal states are two hyperfine states, $\ket{\uparrow}=\ket{F=4,m_F=4}$ and $\ket{\downarrow}=\ket{F=3,m_F=3}$, of the electronic ground state.
Each lattice site can be occupied by at most one atom, due to light-induced collisions in a tight dipole trap potential \cite{Schlosser:2001}.
The atoms are cooled to the motional ground state using microwave sideband cooling and transferred to a given Fock state $\ket{n}$ with microwave sideband transitions \cite{Belmechri2013}.

For this study, we assume a lattice depth $U_0 = \SI{18}{\micro\kelvin}$ (equivalent to $\approx 190\,E_\text{rec}$) at the beginning of the Ramsey sequence, resulting in a trap period $2\pi/\omega \approx \SI{18}{\micro\second}$.
At this trap depth, the probability of an atom tunneling to a neighboring site during the Ramsey sequence is negligible only for the lowest Fock states, $n\leq 5$; for comparison, the tunneling probability is of unit order for $n\geq 8$. 
It should be said that deeper traps hosting a larger number of Fock states with negligible tunneling probability are in general possible and, following Eq.~(\ref{eq:EnergySpacingLattice}), could be used to reduce the effect of the anharmonicity.

\begin{figure}[t]	
	\centering
    \includegraphics[width=\columnwidth]{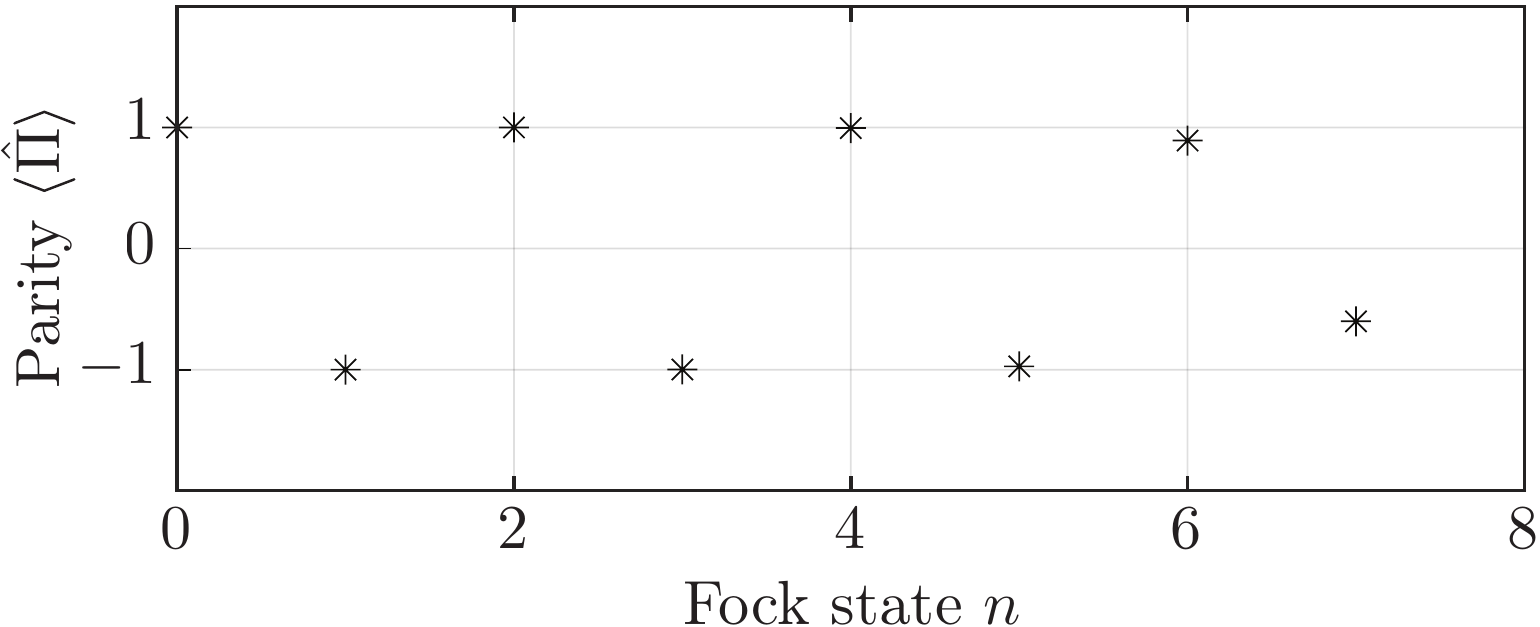}
	\caption{Simulated measurement results of the parity in a lattice potential vs.\ Fock state $n$. The experimental sequence corresponds to measuring the Wigner function at the origin in the phase space.
	}
	\label{fig:Figure3}
\end{figure}

With reference to Figs.~\ref{fig:Wigner_Idea} and \figref{fig:Wigner_EigenenergiesUnbalancedLattice}{a}, the two Ramsey pulses can be realized using copropagating Raman beams \cite{Ness:2021} or microwave radiation.
These methods avoid transferring momentum to the atom that would otherwise alter the motional state and, thus, the reconstructed Wigner function.

For the displacement operation $\hat{D}(x,p)$, one can use an optical conveyor belt \cite{Kuhr:2003}, allowing for precise control of the position of the optical lattice.
In fact, by acting on one of the two laser beams forming the optical lattice, one can suddenly shift the position by $x$ and velocity by $p/m$ of the trap at time $\tau_W$ by simply changing the phase and frequency of the laser beam by $4\pi\,x/\lambda$ and $2 p/(m\,\lambda)$, respectively. This can be done with an acousto-optic modulator. 
In a real experiment, however, one should consider the finite modulation bandwidth of acousto-optic modulators, which causes a smoother variation of position and velocity, instead of a discontinuous change as in the ideal case.
To be realistic, we henceforth assume that the displacement operations vary the trap position smoothly over a time of about $\SI{300}{\nano\second}$, corresponding to about $1/60$-th of the trap period.

\begin{figure*}[t]	
	\centering 
		\includegraphics[width=\linewidth]{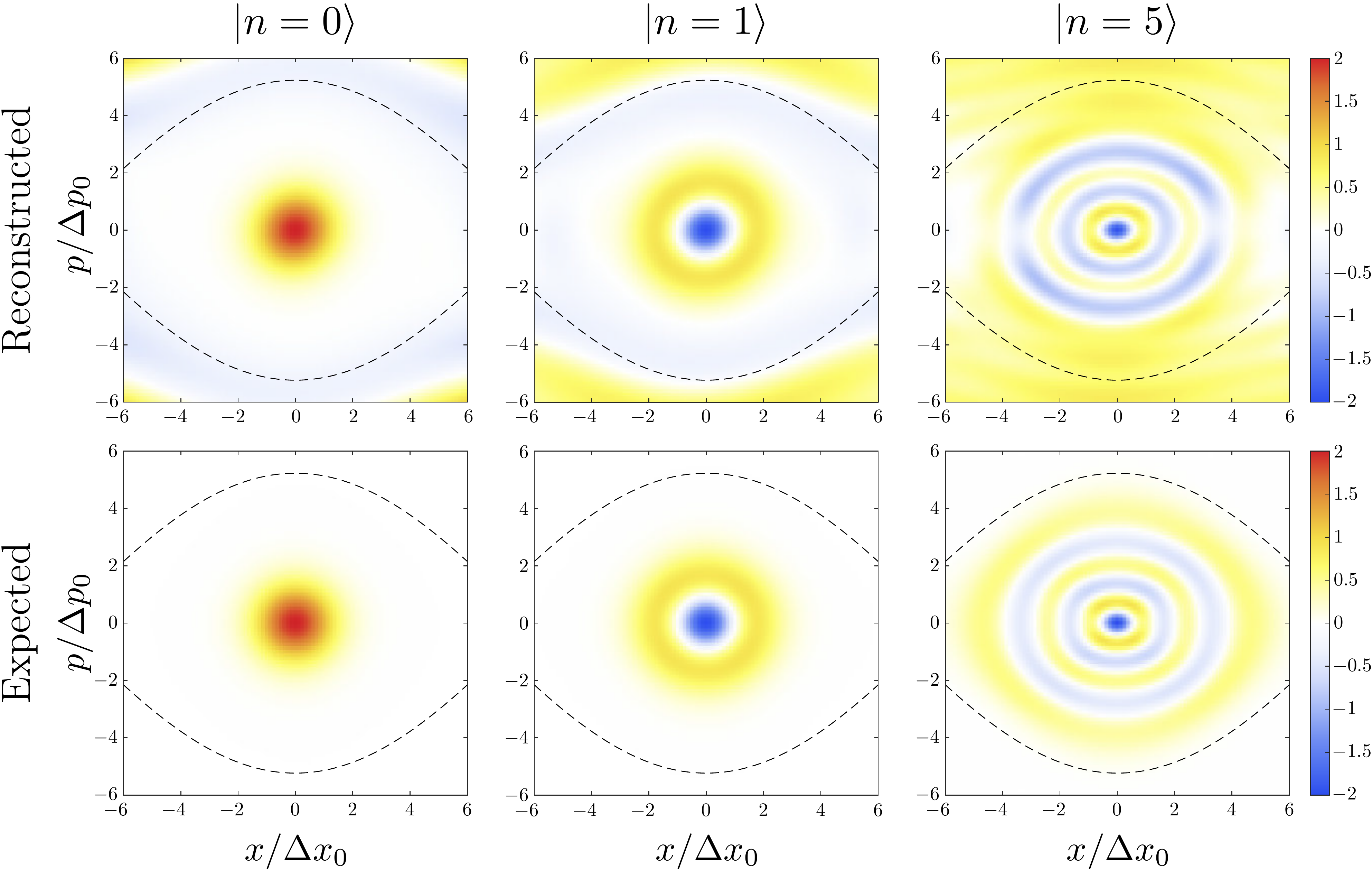}
	\caption{Reconstructed (top row) and expected (bottom row) Wigner functions for the Fock states $n = 0$, $1$ and $5$.
	The displacements in $x$ and $p$ are normalized to the rms widths, $\Delta x_0$ and $\Delta p_0$, of the position and momentum operators evaluated on the ground state, respectively.
	The color scale represents the Wigner function expressed in units of $1/(2\pi\hbar)$.
	The reconstructed Wigner function is obtained from a simulation of the proposed Ramsey scheme, based on a numerical integration of the time-dependent Schr{\"o}dinger equation.
	The dashed lines denote the phase-space points where the Hamiltonian equals zero; outside the of these lines, a corresponding classical particle cannot be trapped.
	Despite the anharmonicity and finite depth of the potential, where only six Fock states experience negligible tunneling and are thus bound to a given lattice site, the reconstructed Wigner function shows remarkable agreement with the expected one.
	Discrepancies are most noticeable for larger $n$ because these states are spatially broader and experience a larger trap anharmonicity.
	The slight rotational distortion for $n=5$ can be ascribed to the finite duration ($\SI{300}{\nano\second}$) of the displacement operator.
		}

	\label{fig:Wigner_CalculatedVsReconstructed}
\end{figure*}

For the parity operation $\hat{\Pi}$, the trap depth must be differentially modulated for the two internal states so as to fulfill the condition in Eq.~(\ref{eq:Wigner_N_PI}).
Such a change can be performed by varying the polarization ellipticity of one of the lattice beams, as demonstrated in Ref.~\cite{Ramola2021a}, by either using an electro-optic modulator or a polarization synthesizer \cite{Robens2018}.
In this study, we consider a smooth variation of the trap depth over a duration of about $\SI{15}{\micro\second}$, as shown in Fig.~\figref{fig:Wigner_EigenenergiesUnbalancedLattice}{a}.
This variation is sufficiently slow that for the relevant, non-tunneling Fock states, $n\leq 5$, the probability of excitation is in the range of a few percent.
Appendix~\ref{app:tau} describes a calibration procedure for differential modulation of the trap depth so as to fulfill Eq.~(\ref{eq:Wigner_N_PI}).
To reduce the effect of anharmonicity, it is advantageous that the trap depth of the two lattice potentials, $U_{0,\uparrow}$ and $U_{0,\downarrow}$, is not only differentially modulated, but also increased at the same time, as shown in Fig.~\figref{fig:Wigner_EigenenergiesUnbalancedLattice}{a}.
Thereby, the wave packet is adiabatically compressed to the trap center, and the effect of anharmonicity is decreased, cf.\ the nonlinear correction terms in Eq.~(\ref{eq:EnergySpacingLattice}), which scale with the powers of $E_\text{rec}/(\hbar\omega)$.
It should be remarked that a state-dependent variation of the trap depth does not require a specific value of the wavelength $\lambda$ but only that at the chosen wavelength, the differential polarizability of the atom does not vanish \cite{Ramola2021a}.

For the state detection, a standard push-out scheme can be used, which relies on a resonant laser beam to remove atoms state dependently.
Alternatively (and non-destructively), state-dependent optical potentials can be used to map atoms depending on their internal state to different positions \cite{Robens:2017,Wu:2019a} that are subsequently detected by high resolution fluorescence imaging \cite{Robens:2017a}.

\section{Simulation results}

We numerically simulate the proposed scheme for measuring $\mathcal{W}(x,p)$ based on the experimental setup described in the previous section.
To this end, we solve the one-dimensional Schrödinger equation for a time-dependent potential using the Strang split-step method~\cite{MacNamara2016}, as done in Ref.~\cite{Lam2021a}.
The numerical simulation allows us to model both the displacement and the parity operations under realistic experimental conditions, which includes considering the full lattice potential and taking into account the finite modulation bandwidth.

For the ideal case of a harmonic oscillator, our simulations show that the parity measurement performed at the origin in phase space yields values in excellent agreement with the theoretical expectation, with relative errors smaller than $10^{-3}$ for the first ten Fock states.
These small residual deviations are ascribed to the creation of differential motional excitations resulting from a small violation of the adiabatic condition.

For a real lattice potential, deviations of the parity measurement are more significant compared to the harmonic oscillator case.
This is because of the nonlinear correction to the energy spectrum that prevents us from fulfilling the parity operator condition in Eq.~(\ref{eq:Wigner_N_PI}) for all Fock states simultaneously.
Figure~\ref{fig:Figure3} shows the simulated parity measurement as a function of Fock state $n$ relative to the origin in phase space.
The results show a near perfect measurement of parity until $n=6$, where we expect the fidelity to begin to degrade due to non-negligible tunneling, thus demonstrating the robustness of the scheme against the anharmonicity of the trapping potential.

Figure~\ref{fig:Wigner_CalculatedVsReconstructed} shows the main results of this work:
The reconstructed Wigner function is plotted for a few representative low-lying Fock states, $n=0$, $1$, and $5$, in units of the rms width of the motional ground state for the position and momentum coordinates, $\Delta x_0 = \sqrt{\hbar/(2m\omega)}$ and $\Delta p_0 = \hbar /(2\Delta x_0)$.
For comparison, we show in the same figure the expected Wigner function, which is derived numerically by applying the original definition \cite{Wigner1932} to the quantum state at time $\tau_W$.
The striking similarity between the reconstructed and expected Wigner functions attests to the  high fidelity of the proposed measurement technique, even for higher Fock states.

\section{Discussion}

The results presented in Fig.~\ref{fig:Wigner_CalculatedVsReconstructed} are remarkable, as we are able to probe states as high as $n=5$ that extend in phase space nearly up to the zero-energy isolines (dashed lines in the figure).
When the displacement operation reaches outside of these isolines, the atom is no longer bound to the original lattice site but is free to move away during the time of the Ramsey interferometer, resulting in a drop of fidelity.

In a lattice potential, the measurement fidelity of the proposed scheme is fundamentally limited by the finite trap depth $U_0$ and finite size of the lattice cell $\lambda/2$.
These two factors are responsible for the anharmonicity of the trap potential.
One possibility for reconstruction error occurs when the atom's kinetic energy increases by more than the trap depth $U_0$, which occurs when the radius $p/\Delta p_0$ (and consequently $r/\Delta r_0$, owing to the dynamics in the trap) is comparable or exceeds $\pm\sqrt{2}(U_0/E_\text{rec})^{1/4} \approx \pm 5$.
The other possibility for reconstruction error arises when the shifted atom is subject to a highly anharmonic trap with inverted potential curvature.
This is the case when the shift is larger than $\pm \lambda/8$, corresponding to a radius $x/\Delta x_0$ (and consequently $p/\Delta p_0$ owing to the dynamics in the trap) comparable to or exceeding $\pm (U_0/E_\text{rec})^{1/4}\pi/(2\sqrt{2}) \approx \pm 4$.
An even larger shift exceeding $\pm \lambda/2$ results in the atom being trapped in the potential of the adjacent lattice site.
Hence, this analysis shows that for an optical lattice potential, the two limits set by the finite trap depth and finite lattice constant produce similar bounds in phase space.
Another limiting factor is the spatial homogeneity of the trap potential.
We recall from Eq.~\eqref{eq:DeltaE_HO} that the differential phase acquired between the two internal states for a given Fock state, $\Phi(n)$, depends both on the differential trap frequency $\Delta\omega$ and the differential energy offset $\Delta E_0$.
The energy offset, being dependent on the trap depth, makes our proposed method sensitive to spatial variations of the lattice depth.
Moreover, it should be noted that $\Delta E_0$ is a much larger quantity than $\hbar\Delta \omega$, since it scales with the trap depth, rather than the trap frequency.
Thus, when an ensemble of atoms sparsely fills the lattice, a spatial variation of the trap depth produces an inhomogeneous broadening of the phase $\Phi_0$ of the Ramsey fringe, which cannot be fully compensated by suitably choosing the relative phase between the two $\pi/2$ pulses.
As a result, the Ramsey fringe contrast decreases and so does the visibility of the reconstructed Wigner function.
This detrimental effect can be avoided by probing only few atoms at a time in a small homogeneous region of the lattice.

For similar reasons, it is important that the atoms are prepared in the motional ground state (or any other well defined state) in the two other spatial dimensions that are not probed. Otherwise, an inhomogeneous spread of motional energies (e.g., due to thermal motion in these dimensions) would smear out $\Phi_0$ and consequently reduce the visibility of the measured Wigner function.

Our proposed technique is not limited to optical lattices but can also be applied to optical tweezers, provided that sufficiently fast, coherent transport can be realized, e.g. via electro-optic or acousto-optic deflectors~\cite{Barredo, Kjaergaard, Henderson_2009}. It is challenging to realize high trapping frequencies in optical tweezers, but a similar degree of common-mode suppression of the anharmonicity can be expected because the energy spacing is also insensitive to the trap frequency up to first order in perturbation theory, as we show in Appendix~\ref{app:tweezer}.
Moreover, in this case, optical tweezers have only one potential minimum and the potential goes to zero at infinity. Thus these potentials do not have to deal with the effects of adjacent site tunneling.
In this case, however, the atoms experience a much weaker confinement and eventually will no longer be trapped due to, e.g., gravity.

Finally, the proposed reconstruction technique is not limited to any specific atomic species or trapping wavelength. It is sufficient that the qubit state experiences a vectorial differential light shift when the polarization ellipticity is varied, as discussed in Ref.~\cite{Ramola2021a}.

\section{Conclusion}

We have presented and numerically investigated a novel scheme to directly measure the Wigner function of neutral atoms trapped in an anharmonic potential.
We have discussed the measurement technique using an optical lattice potential as an example, and we are confident that the proposed scheme can also be applied to neutral atoms trapped in optical tweezers.
Our simulations suggest that the proposed implementation of the parity operation is sufficiently robust to measure the Wigner function of the lowest six Fock states in a trap where only these six states have a negligible probability of tunneling to an adjacent lattice site.
The temporal resolution is technically limited by how fast the displacement operation can be implemented.
Technical challenges notwithstanding, the measurement fidelity is ultimately limited by nonlinearities that arise from the anharmonicity of the potential and, in the case of a lattice potential, atom tunneling into adjacent sites.

The proposed scheme also appears to be experimentally feasible.
In previous work, we have demonstrated fast displacement of lattice potentials for fast coherent transport \cite{Lam2021a}, the application of Ramsey interferometry to track the dynamics of a matter wave excitation \cite{Ness:2021}, and used a differential modulation of the trap depth to interferometrically measure the optical potential landscape with high spatial resolution \cite{Ramola2021a}.
In the same publications \cite{Lam2021a,Ness:2021,Ramola2021a}, we have shown that the coherence time of the atoms remains unaffected despite a state-dependent modulation of the lattice position and depth, which is implicitly assumed in the simulations presented here.
Thus, we hope that the proposed method, when experimentally realized, can extend the toolbox available for precise quantum state reconstruction and can inspire other applications in the field of quantum metrology.

\section{Acknowledgments} This research was supported by the Collaborative Research Center SFB/TR 185 OSCAR of the German Research Foundation.

\appendix

\makeatletter
\def\section#1{\old@Section@Cmd{#1}}
\makeatother

\begin{figure}[b]	
	\centering 
	\includegraphics[width=\columnwidth]{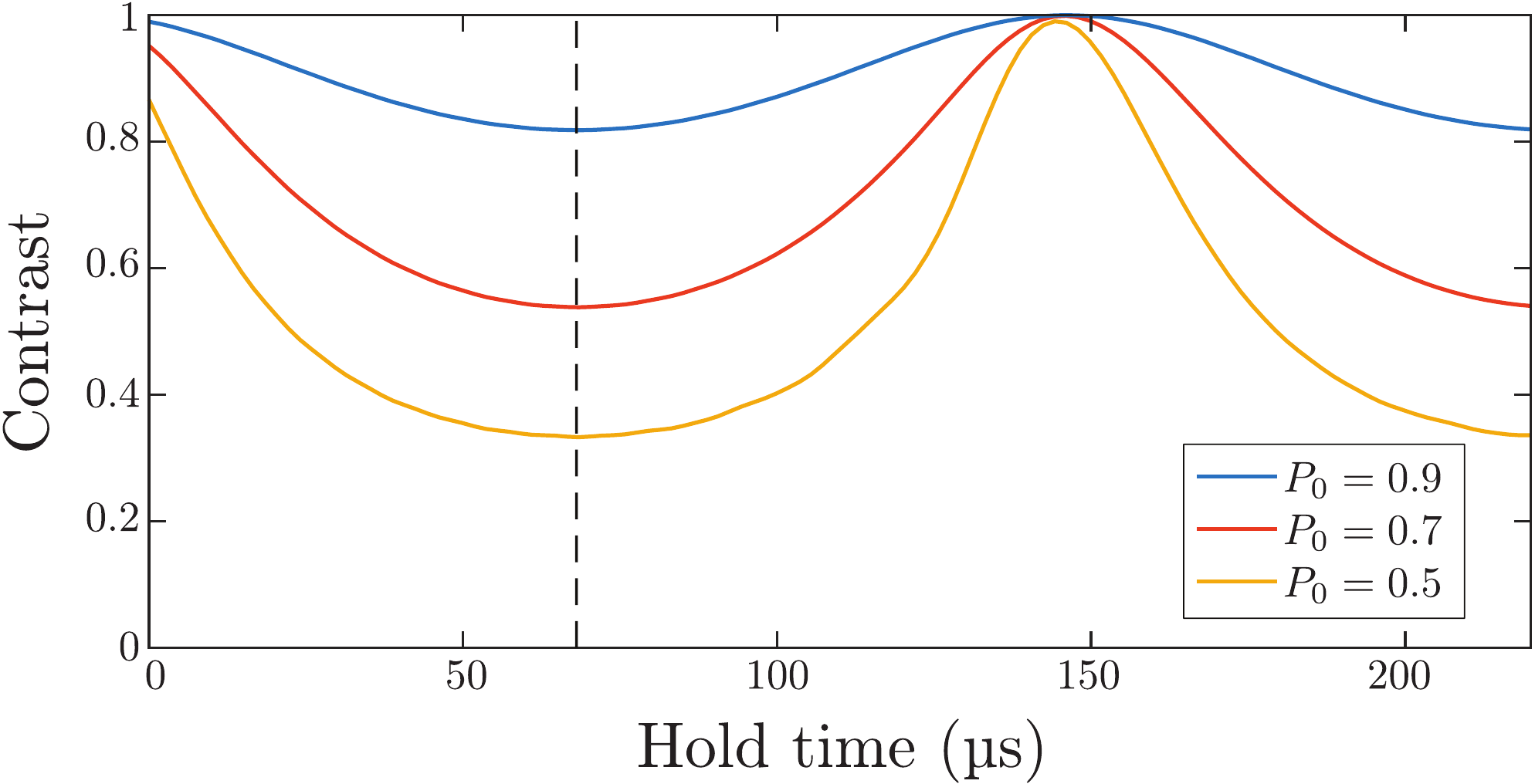}
	\caption{Plot of the Ramsey fringe contrast as a function of the hold time at the maximum differential trap frequency, with $P_0$ denoting the ground state fraction of a thermal state.
	The same sequence as in Fig.~\figref{fig:Wigner_EigenenergiesUnbalancedLattice}{a} is applied with the displacement set to zero.
	The optimal hold time is chosen as the time when the contrast is minimized; this occurs when the phase difference between states of opposite parity is $\pi$.
	We note that the anharmonicity of the trap hinders a complete revival in the case of small $P_0$ because the number of Fock states occupied is higher and thus the effect of the trap anharmonicity is greater.
	}
	\label{fig:calibration}
\end{figure}

\section{Calibration procedure}
\label{app:tau}

The differential modulation of the trap depth must be calibrated in order to satisfy Eq.~(\ref{eq:Wigner_N_PI}). 
To that end, one can apply to a thermal state the same sequence described in Fig.~\figref{fig:Wigner_EigenenergiesUnbalancedLattice}{a}.
The displacement can simply be assumed to be zero, although other values can be chosen since this calibration is largely independent of the phase space point chosen.
To perform this calibration, one scans the probing time in the unbalanced trap (though, the scan parameter could also be the degree of trap imbalance at a fixed probing time) and records the Ramsey fringe contrast.
Figure~\ref{fig:calibration} shows the simulated Ramsey contrast as a function of the hold time, with clearly visible collapses and revivals;
the dashed vertical line in the figure indicates the time of the first collapse.
The amplitude of the collapses increases as the probability $P_0$ of occupying the motional ground state decreases, i.e., when more Fock states are occupied by a given thermal state.
Importantly, a collapse occurs when the different occupied Fock states have maximally dephased in the Ramsey sequence.
This is the case when the condition in Eq.~(\ref{eq:Wigner_N_PI}) is fulfilled.

It might be beneficial to repeat this calibration procedure before probing points in phase space with significantly different $p$-values.
This is due to the facft that, when the atoms are dragged by the trap at the velocity $p/m$ for the duration of the Ramsey interferometer, they experience the spatial inhomogeneity of the laser beams forming the lattice.
Without recalibration, the resulting variation of the trap depth can negatively affect the condition in Eq.~(\ref{eq:Wigner_N_PI}), as well as the condition for the cancellation of the phase shift $\Phi_0$.

\section{Optical tweezers}
\label{app:tweezer}

The proposed scheme can be readily adapted to measure the Wigner function of atoms trapped in an optical tweezer produced by a tightly focused laser beam of waist $w$. The optical tweezer potential in the transverse direction reads
$U(x)=U_0\,\exp(-2x^2/w^2)$, where $U_0$ is the trap depth.
In first order perturbation theory, the energy spectrum resulting from the tweezer potential is
\begin{equation}
E_n = E^\mathrm{(HO)}_n-\frac{3[{2n(n+1)+1}]\hbar^2}{8\hspace{0.3pt}m\hspace{0.3pt}w^2} + \hbar\omega\,\mathcal{O}\left(\frac{\hbar^2}{m\hspace{0.3pt}w^2U_0}\right).
\label{eq:EnergySpacingHarmonicTrap}
\end{equation}
As with the spectrum of an optical lattice [cf.\ Eq.~(\ref{eq:EnergySpacingLattice})], the spectrum of an optical tweezer shows that the anharmonic contribution to the energy spacing is, up to first order, independent of the trap frequency $\omega$.
We can therefore expect a similar common-mode suppression of the nonlinear contribution for tweezers as we discussed for the case of an optical lattice.
In addition, single tweezers, in contrast to lattice potentials, have the advantage that atoms cannot tunnel to a neighboring trap.
\newpage


\begin{thebibliography}{45}%
\makeatletter
\providecommand \@ifxundefined [1]{%
 \@ifx{#1\undefined}
}%
\providecommand \@ifnum [1]{%
 \ifnum #1\expandafter \@firstoftwo
 \else \expandafter \@secondoftwo
 \fi
}%
\providecommand \@ifx [1]{%
 \ifx #1\expandafter \@firstoftwo
 \else \expandafter \@secondoftwo
 \fi
}%
\providecommand \natexlab [1]{#1}%
\providecommand \enquote  [1]{``#1''}%
\providecommand \bibnamefont  [1]{#1}%
\providecommand \bibfnamefont [1]{#1}%
\providecommand \citenamefont [1]{#1}%
\providecommand \href@noop [0]{\@secondoftwo}%
\providecommand \href [0]{\begingroup \@sanitize@url \@href}%
\providecommand \@href[1]{\@@startlink{#1}\@@href}%
\providecommand \@@href[1]{\endgroup#1\@@endlink}%
\providecommand \@sanitize@url [0]{\catcode `\\12\catcode `\$12\catcode
  `\&12\catcode `\#12\catcode `\^12\catcode `\_12\catcode `\%12\relax}%
\providecommand \@@startlink[1]{}%
\providecommand \@@endlink[0]{}%
\providecommand \url  [0]{\begingroup\@sanitize@url \@url }%
\providecommand \@url [1]{\endgroup\@href {#1}{\urlprefix }}%
\providecommand \urlprefix  [0]{URL }%
\providecommand \Eprint [0]{\href }%
\providecommand \doibase [0]{https://doi.org/}%
\providecommand \selectlanguage [0]{\@gobble}%
\providecommand \bibinfo  [0]{\@secondoftwo}%
\providecommand \bibfield  [0]{\@secondoftwo}%
\providecommand \translation [1]{[#1]}%
\providecommand \BibitemOpen [0]{}%
\providecommand \bibitemStop [0]{}%
\providecommand \bibitemNoStop [0]{.\EOS\space}%
\providecommand \EOS [0]{\spacefactor3000\relax}%
\providecommand \BibitemShut  [1]{\csname bibitem#1\endcsname}%
\let\auto@bib@innerbib\@empty
\bibitem [{\citenamefont {Wigner}(1932)}]{Wigner1932}%
  \BibitemOpen
  \bibfield  {author} {\bibinfo {author} {\bibfnamefont {E.}~\bibnamefont
  {Wigner}},\ }\bibfield  {title} {\bibinfo {title} {{On the Quantum Correction
  For Thermodynamic Equilibrium}},\ }\href
  {https://doi.org/10.1103/PhysRev.40.749} {\bibfield  {journal} {\bibinfo
  {journal} {Phys. Rev.}\ }\textbf {\bibinfo {volume} {40}},\ \bibinfo {pages}
  {749} (\bibinfo {year} {1932})}\BibitemShut {NoStop}%
\bibitem [{\citenamefont {Groenewold}(1946)}]{Groenewold1946}%
  \BibitemOpen
  \bibfield  {author} {\bibinfo {author} {\bibfnamefont {H.~J.}\ \bibnamefont
  {Groenewold}},\ }\bibfield  {title} {\bibinfo {title} {{On the principles of
  elementary quantum mechanics}},\ }\href
  {https://doi.org/10.1016/S0031-8914(46)80059-4} {\bibfield  {journal}
  {\bibinfo  {journal} {Physica}\ }\textbf {\bibinfo {volume} {12}},\ \bibinfo
  {pages} {405} (\bibinfo {year} {1946})}\BibitemShut {NoStop}%
\bibitem [{\citenamefont {Moyal}\ and\ \citenamefont
  {Bartlett}(1949)}]{Moyal1949}%
  \BibitemOpen
  \bibfield  {author} {\bibinfo {author} {\bibfnamefont {J.~E.}\ \bibnamefont
  {Moyal}}\ and\ \bibinfo {author} {\bibfnamefont {M.~S.}\ \bibnamefont
  {Bartlett}},\ }\bibfield  {title} {\bibinfo {title} {{Quantum mechanics as a
  statistical theory}},\ }\href {https://doi.org/10.1017/S0305004100000487}
  {\bibfield  {journal} {\bibinfo  {journal} {Proc. Cambridge. Philos. Soc.}\
  }\textbf {\bibinfo {volume} {45}},\ \bibinfo {pages} {99} (\bibinfo {year}
  {1949})}\BibitemShut {NoStop}%
\bibitem [{\citenamefont {Hillery}\ \emph {et~al.}(1984)\citenamefont
  {Hillery}, \citenamefont {O'Connell}, \citenamefont {Scully},\ and\
  \citenamefont {Wigner}}]{Hillery1984}%
  \BibitemOpen
  \bibfield  {author} {\bibinfo {author} {\bibfnamefont {M.}~\bibnamefont
  {Hillery}}, \bibinfo {author} {\bibfnamefont {R.~F.}\ \bibnamefont
  {O'Connell}}, \bibinfo {author} {\bibfnamefont {M.~O.}\ \bibnamefont
  {Scully}},\ and\ \bibinfo {author} {\bibfnamefont {E.~P.}\ \bibnamefont
  {Wigner}},\ }\bibfield  {title} {\bibinfo {title} {{Distribution functions in
  physics: Fundamentals}},\ }\href
  {https://doi.org/10.1016/0370-1573(84)90160-1} {\bibfield  {journal}
  {\bibinfo  {journal} {Phys. Rep.}\ }\textbf {\bibinfo {volume} {106}},\
  \bibinfo {pages} {121} (\bibinfo {year} {1984})}\BibitemShut {NoStop}%
\bibitem [{\citenamefont {Zachos}\ \emph {et~al.}(2005)\citenamefont {Zachos},
  \citenamefont {Fairlie},\ and\ \citenamefont {Curtright}}]{Zachos2005}%
  \BibitemOpen
  \bibfield  {author} {\bibinfo {author} {\bibfnamefont {C.~K.}\ \bibnamefont
  {Zachos}}, \bibinfo {author} {\bibfnamefont {D.~B.}\ \bibnamefont
  {Fairlie}},\ and\ \bibinfo {author} {\bibfnamefont {T.~L.}\ \bibnamefont
  {Curtright}},\ }\href {https://doi.org/10.1142/5287} {\emph {\bibinfo {title}
  {{Quantum Mechanics in Phase Space: an Overview with Selected Papers}}}},\
  edited by\ \bibinfo {editor} {\bibfnamefont {C.~K.}\ \bibnamefont {Zachos}},
  \bibinfo {editor} {\bibfnamefont {D.~B.}\ \bibnamefont {Fairlie}},\ and\
  \bibinfo {editor} {\bibfnamefont {T.~L.}\ \bibnamefont {Curtright}}\
  (\bibinfo  {publisher} {World Scientific Publishing},\ \bibinfo {year}
  {2005})\BibitemShut {NoStop}%
\bibitem [{\citenamefont {Hudson}(1974)}]{Hudson_1974}%
  \BibitemOpen
  \bibfield  {author} {\bibinfo {author} {\bibfnamefont {R.}~\bibnamefont
  {Hudson}},\ }\bibfield  {title} {\bibinfo {title} {{When is the Wigner
  quasi-probability density non-negative?}},\ }\href
  {https://doi.org/https://doi.org/10.1016/0034-4877(74)90007-X} {\bibfield
  {journal} {\bibinfo  {journal} {Rep. Math. Phys.}\ }\textbf {\bibinfo
  {volume} {6}},\ \bibinfo {pages} {249} (\bibinfo {year} {1974})}\BibitemShut
  {NoStop}%
\bibitem [{\citenamefont {Kenfack}\ and\ \citenamefont
  {Zyczkowski}(2004)}]{Kenfack_2004}%
  \BibitemOpen
  \bibfield  {author} {\bibinfo {author} {\bibfnamefont {A.}~\bibnamefont
  {Kenfack}}\ and\ \bibinfo {author} {\bibfnamefont {K.}~\bibnamefont
  {Zyczkowski}},\ }\bibfield  {title} {\bibinfo {title} {{Negativity of the
  Wigner function as an indicator of non-classicality}},\ }\href
  {https://doi.org/10.1088/1464-4266/6/10/003} {\bibfield  {journal} {\bibinfo
  {journal} {J. Opt. B: Quantum. Semiclassical. Opt.}\ }\textbf {\bibinfo
  {volume} {6}},\ \bibinfo {pages} {396} (\bibinfo {year} {2004})}\BibitemShut
  {NoStop}%
\bibitem [{\citenamefont {Del{\'e}glise}\ \emph {et~al.}(2008)\citenamefont
  {Del{\'e}glise}, \citenamefont {Dotsenko}, \citenamefont {Sayrin},
  \citenamefont {Bernu}, \citenamefont {Brune}, \citenamefont {Raimond},\ and\
  \citenamefont {Haroche}}]{Deleglise2008}%
  \BibitemOpen
  \bibfield  {author} {\bibinfo {author} {\bibfnamefont {S.}~\bibnamefont
  {Del{\'e}glise}}, \bibinfo {author} {\bibfnamefont {I.}~\bibnamefont
  {Dotsenko}}, \bibinfo {author} {\bibfnamefont {C.}~\bibnamefont {Sayrin}},
  \bibinfo {author} {\bibfnamefont {J.}~\bibnamefont {Bernu}}, \bibinfo
  {author} {\bibfnamefont {M.}~\bibnamefont {Brune}}, \bibinfo {author}
  {\bibfnamefont {J.-M.}\ \bibnamefont {Raimond}},\ and\ \bibinfo {author}
  {\bibfnamefont {S.}~\bibnamefont {Haroche}},\ }\bibfield  {title} {\bibinfo
  {title} {{Reconstruction of non-classical cavity field states with snapshots
  of their decoherence}},\ }\href {https://doi.org/10.1038/nature07288}
  {\bibfield  {journal} {\bibinfo  {journal} {Nature}\ }\textbf {\bibinfo
  {volume} {455}},\ \bibinfo {pages} {510} (\bibinfo {year}
  {2008})}\BibitemShut {NoStop}%
\bibitem [{\citenamefont {Weinbub}\ and\ \citenamefont
  {Ferry}(2018)}]{Weinbub2018}%
  \BibitemOpen
  \bibfield  {author} {\bibinfo {author} {\bibfnamefont {J.}~\bibnamefont
  {Weinbub}}\ and\ \bibinfo {author} {\bibfnamefont {D.~K.}\ \bibnamefont
  {Ferry}},\ }\bibfield  {title} {\bibinfo {title} {{Recent advances in Wigner
  function approaches}},\ }\href {https://doi.org/10.1063/1.5046663} {\bibfield
   {journal} {\bibinfo  {journal} {Appl. Phys. Rev.}\ }\textbf {\bibinfo
  {volume} {5}},\ \bibinfo {pages} {041104} (\bibinfo {year}
  {2018})}\BibitemShut {NoStop}%
\bibitem [{\citenamefont {Leibfried}\ \emph {et~al.}(1997)\citenamefont
  {Leibfried}, \citenamefont {Meekhof}, \citenamefont {Monroe}, \citenamefont
  {King}, \citenamefont {Itano},\ and\ \citenamefont
  {Wineland}}]{Leibfried1997}%
  \BibitemOpen
  \bibfield  {author} {\bibinfo {author} {\bibfnamefont {D.}~\bibnamefont
  {Leibfried}}, \bibinfo {author} {\bibfnamefont {D.~M.}\ \bibnamefont
  {Meekhof}}, \bibinfo {author} {\bibfnamefont {C.}~\bibnamefont {Monroe}},
  \bibinfo {author} {\bibfnamefont {B.~E.}\ \bibnamefont {King}}, \bibinfo
  {author} {\bibfnamefont {W.~M.}\ \bibnamefont {Itano}},\ and\ \bibinfo
  {author} {\bibfnamefont {D.~J.}\ \bibnamefont {Wineland}},\ }\bibfield
  {title} {\bibinfo {title} {{Experimental preparation and measurement of
  quantum states of motion of a trapped atom}},\ }\href
  {https://doi.org/10.1080/09500349708231896} {\bibfield  {journal} {\bibinfo
  {journal} {J. Mod. Opt.}\ }\textbf {\bibinfo {volume} {44}},\ \bibinfo
  {pages} {2485} (\bibinfo {year} {1997})}\BibitemShut {NoStop}%
\bibitem [{\citenamefont {Lv}\ \emph {et~al.}(2017)\citenamefont {Lv},
  \citenamefont {An}, \citenamefont {Um}, \citenamefont {Zhang}, \citenamefont
  {Zhang}, \citenamefont {Kim},\ and\ \citenamefont {Kim}}]{Lv:2017}%
  \BibitemOpen
  \bibfield  {author} {\bibinfo {author} {\bibfnamefont {D.}~\bibnamefont
  {Lv}}, \bibinfo {author} {\bibfnamefont {S.}~\bibnamefont {An}}, \bibinfo
  {author} {\bibfnamefont {M.}~\bibnamefont {Um}}, \bibinfo {author}
  {\bibfnamefont {J.}~\bibnamefont {Zhang}}, \bibinfo {author} {\bibfnamefont
  {J.-N.}\ \bibnamefont {Zhang}}, \bibinfo {author} {\bibfnamefont {M.~S.}\
  \bibnamefont {Kim}},\ and\ \bibinfo {author} {\bibfnamefont {K.}~\bibnamefont
  {Kim}},\ }\bibfield  {title} {\bibinfo {title} {{Reconstruction of the
  Jaynes-Cummings field state of ionic motion in a harmonic trap}},\ }\href
  {https://doi.org/10.1103/PhysRevA.95.043813} {\bibfield  {journal} {\bibinfo
  {journal} {Phys. Rev. A}\ }\textbf {\bibinfo {volume} {95}},\ \bibinfo
  {pages} {043813} (\bibinfo {year} {2017})}\BibitemShut {NoStop}%
\bibitem [{\citenamefont {Poyatos}\ \emph {et~al.}(1996)\citenamefont
  {Poyatos}, \citenamefont {Walser}, \citenamefont {Cirac}, \citenamefont
  {Zoller},\ and\ \citenamefont {Blatt}}]{Poyatos1966}%
  \BibitemOpen
  \bibfield  {author} {\bibinfo {author} {\bibfnamefont {J.~F.}\ \bibnamefont
  {Poyatos}}, \bibinfo {author} {\bibfnamefont {R.}~\bibnamefont {Walser}},
  \bibinfo {author} {\bibfnamefont {J.~I.}\ \bibnamefont {Cirac}}, \bibinfo
  {author} {\bibfnamefont {P.}~\bibnamefont {Zoller}},\ and\ \bibinfo {author}
  {\bibfnamefont {R.}~\bibnamefont {Blatt}},\ }\bibfield  {title} {\bibinfo
  {title} {{Motion tomography of a single trapped ion}},\ }\href
  {https://doi.org/10.1103/PhysRevA.53.R1966} {\bibfield  {journal} {\bibinfo
  {journal} {Phys. Rev. A}\ }\textbf {\bibinfo {volume} {53}},\ \bibinfo
  {pages} {R1966} (\bibinfo {year} {1996})}\BibitemShut {NoStop}%
\bibitem [{\citenamefont {Lvovsky}\ \emph {et~al.}(2001)\citenamefont
  {Lvovsky}, \citenamefont {Hansen}, \citenamefont {Aichele}, \citenamefont
  {Benson}, \citenamefont {Mlynek},\ and\ \citenamefont
  {Schiller}}]{Lvovsky:2001}%
  \BibitemOpen
  \bibfield  {author} {\bibinfo {author} {\bibfnamefont {A.~I.}\ \bibnamefont
  {Lvovsky}}, \bibinfo {author} {\bibfnamefont {H.}~\bibnamefont {Hansen}},
  \bibinfo {author} {\bibfnamefont {T.}~\bibnamefont {Aichele}}, \bibinfo
  {author} {\bibfnamefont {O.}~\bibnamefont {Benson}}, \bibinfo {author}
  {\bibfnamefont {J.}~\bibnamefont {Mlynek}},\ and\ \bibinfo {author}
  {\bibfnamefont {S.}~\bibnamefont {Schiller}},\ }\bibfield  {title} {\bibinfo
  {title} {{Quantum State Reconstruction of the Single-Photon Fock State}},\
  }\href {https://doi.org/10.1103/PhysRevLett.87.050402} {\bibfield  {journal}
  {\bibinfo  {journal} {Phys. Rev. Lett.}\ }\textbf {\bibinfo {volume} {87}},\
  \bibinfo {pages} {050402} (\bibinfo {year} {2001})}\BibitemShut {NoStop}%
\bibitem [{\citenamefont {Fl{\"u}hmann}\ and\ \citenamefont
  {Home}(2020)}]{Fluhmann:2020}%
  \BibitemOpen
  \bibfield  {author} {\bibinfo {author} {\bibfnamefont {C.}~\bibnamefont
  {Fl{\"u}hmann}}\ and\ \bibinfo {author} {\bibfnamefont {J.~P.}\ \bibnamefont
  {Home}},\ }\bibfield  {title} {\bibinfo {title} {{Direct
  Characteristic-Function Tomography of Quantum States of the Trapped-Ion
  Motional Oscillator}},\ }\href
  {https://doi.org/10.1103/PhysRevLett.125.043602} {\bibfield  {journal}
  {\bibinfo  {journal} {Phys. Rev. Lett.}\ }\textbf {\bibinfo {volume} {125}},\
  \bibinfo {pages} {043602} (\bibinfo {year} {2020})}\BibitemShut {NoStop}%
\bibitem [{\citenamefont {Cahill}\ and\ \citenamefont
  {Glauber}(1969{\natexlab{a}})}]{Cahill1969a}%
  \BibitemOpen
  \bibfield  {author} {\bibinfo {author} {\bibfnamefont {K.~E.}\ \bibnamefont
  {Cahill}}\ and\ \bibinfo {author} {\bibfnamefont {R.~J.}\ \bibnamefont
  {Glauber}},\ }\bibfield  {title} {\bibinfo {title} {{Ordered Expansions in
  Boson Amplitude Operators}},\ }\href
  {https://doi.org/10.1103/PhysRev.177.1857} {\bibfield  {journal} {\bibinfo
  {journal} {Phys. Rev.}\ }\textbf {\bibinfo {volume} {177}},\ \bibinfo {pages}
  {1857} (\bibinfo {year} {1969}{\natexlab{a}})}\BibitemShut {NoStop}%
\bibitem [{\citenamefont {Cahill}\ and\ \citenamefont
  {Glauber}(1969{\natexlab{b}})}]{Cahill1969}%
  \BibitemOpen
  \bibfield  {author} {\bibinfo {author} {\bibfnamefont {K.~E.}\ \bibnamefont
  {Cahill}}\ and\ \bibinfo {author} {\bibfnamefont {R.~J.}\ \bibnamefont
  {Glauber}},\ }\bibfield  {title} {\bibinfo {title} {{Density Operators and
  Quasiprobability Distributions}},\ }\href
  {https://doi.org/10.1103/PhysRev.177.1882} {\bibfield  {journal} {\bibinfo
  {journal} {Phys. Rev.}\ }\textbf {\bibinfo {volume} {177}},\ \bibinfo {pages}
  {1882} (\bibinfo {year} {1969}{\natexlab{b}})}\BibitemShut {NoStop}%
\bibitem [{\citenamefont {Grossmann}(1976)}]{Grossmann1976}%
  \BibitemOpen
  \bibfield  {author} {\bibinfo {author} {\bibfnamefont {A.}~\bibnamefont
  {Grossmann}},\ }\bibfield  {title} {\bibinfo {title} {{Parity operator and
  quantization of $\delta$-functions}},\ }\href
  {https://doi.org/10.1007/BF01617867} {\bibfield  {journal} {\bibinfo
  {journal} {Commun. Math. Phys.}\ }\textbf {\bibinfo {volume} {48}},\ \bibinfo
  {pages} {191} (\bibinfo {year} {1976})}\BibitemShut {NoStop}%
\bibitem [{\citenamefont {Royer}(1977)}]{Royer1977}%
  \BibitemOpen
  \bibfield  {author} {\bibinfo {author} {\bibfnamefont {A.}~\bibnamefont
  {Royer}},\ }\bibfield  {title} {\bibinfo {title} {{Wigner function as the
  expectation value of a parity operator}},\ }\href
  {https://doi.org/10.1103/PhysRevA.15.449} {\bibfield  {journal} {\bibinfo
  {journal} {Phys. Rev. A}\ }\textbf {\bibinfo {volume} {15}},\ \bibinfo
  {pages} {449} (\bibinfo {year} {1977})}\BibitemShut {NoStop}%
\bibitem [{\citenamefont {Birrittella}\ \emph {et~al.}(2021)\citenamefont
  {Birrittella}, \citenamefont {Alsing},\ and\ \citenamefont
  {Gerry}}]{Birrittella2021}%
  \BibitemOpen
  \bibfield  {author} {\bibinfo {author} {\bibfnamefont {R.~J.}\ \bibnamefont
  {Birrittella}}, \bibinfo {author} {\bibfnamefont {P.~M.}\ \bibnamefont
  {Alsing}},\ and\ \bibinfo {author} {\bibfnamefont {C.~C.}\ \bibnamefont
  {Gerry}},\ }\bibfield  {title} {\bibinfo {title} {{The parity operator:
  Applications in quantum metrology}},\ }\href
  {https://doi.org/10.1116/5.0026148} {\bibfield  {journal} {\bibinfo
  {journal} {AVS Quantum Science}\ }\textbf {\bibinfo {volume} {3}},\ \bibinfo
  {pages} {014701} (\bibinfo {year} {2021})}\BibitemShut {NoStop}%
\bibitem [{\citenamefont {Leibfried}\ \emph {et~al.}(1996)\citenamefont
  {Leibfried}, \citenamefont {Meekhof}, \citenamefont {King}, \citenamefont
  {Monroe}, \citenamefont {Itano},\ and\ \citenamefont
  {Wineland}}]{Leibfried:1996}%
  \BibitemOpen
  \bibfield  {author} {\bibinfo {author} {\bibfnamefont {D.}~\bibnamefont
  {Leibfried}}, \bibinfo {author} {\bibfnamefont {D.~M.}\ \bibnamefont
  {Meekhof}}, \bibinfo {author} {\bibfnamefont {B.~E.}\ \bibnamefont {King}},
  \bibinfo {author} {\bibfnamefont {C.}~\bibnamefont {Monroe}}, \bibinfo
  {author} {\bibfnamefont {W.~M.}\ \bibnamefont {Itano}},\ and\ \bibinfo
  {author} {\bibfnamefont {D.~J.}\ \bibnamefont {Wineland}},\ }\bibfield
  {title} {\bibinfo {title} {{Experimental Determination of the Motional
  Quantum State of a Trapped Atom}},\ }\href
  {https://doi.org/10.1103/PhysRevLett.77.4281} {\bibfield  {journal} {\bibinfo
   {journal} {Phys. Rev. Lett.}\ }\textbf {\bibinfo {volume} {77}},\ \bibinfo
  {pages} {4281} (\bibinfo {year} {1996})}\BibitemShut {NoStop}%
\bibitem [{\citenamefont {Banaszek}\ \emph {et~al.}(1999)\citenamefont
  {Banaszek}, \citenamefont {Radzewicz}, \citenamefont {W{\'o}dkiewicz},\ and\
  \citenamefont {Krasi{\'n}ski}}]{Banaszek1999}%
  \BibitemOpen
  \bibfield  {author} {\bibinfo {author} {\bibfnamefont {K.}~\bibnamefont
  {Banaszek}}, \bibinfo {author} {\bibfnamefont {C.}~\bibnamefont {Radzewicz}},
  \bibinfo {author} {\bibfnamefont {K.}~\bibnamefont {W{\'o}dkiewicz}},\ and\
  \bibinfo {author} {\bibfnamefont {J.~S.}\ \bibnamefont {Krasi{\'n}ski}},\
  }\bibfield  {title} {\bibinfo {title} {{Direct measurement of the Wigner
  function by photon counting}},\ }\href
  {https://doi.org/10.1103/PhysRevA.60.674} {\bibfield  {journal} {\bibinfo
  {journal} {Phys. Rev. A}\ }\textbf {\bibinfo {volume} {60}},\ \bibinfo
  {pages} {674} (\bibinfo {year} {1999})}\BibitemShut {NoStop}%
\bibitem [{\citenamefont {Lutterbach}\ and\ \citenamefont
  {Davidovich}(1997)}]{Lutterbach1997}%
  \BibitemOpen
  \bibfield  {author} {\bibinfo {author} {\bibfnamefont {L.~G.}\ \bibnamefont
  {Lutterbach}}\ and\ \bibinfo {author} {\bibfnamefont {L.}~\bibnamefont
  {Davidovich}},\ }\bibfield  {title} {\bibinfo {title} {{Method for Direct
  Measurement of the Wigner Function in Cavity QED and Ion Traps}},\ }\href
  {https://doi.org/10.1103/PhysRevLett.78.2547} {\bibfield  {journal} {\bibinfo
   {journal} {Phys. Rev. Lett.}\ }\textbf {\bibinfo {volume} {78}},\ \bibinfo
  {pages} {2547} (\bibinfo {year} {1997})}\BibitemShut {NoStop}%
\bibitem [{\citenamefont {Nogues}\ \emph {et~al.}(2000)\citenamefont {Nogues},
  \citenamefont {Rauschenbeutel}, \citenamefont {Osnaghi}, \citenamefont
  {Bertet}, \citenamefont {Brune}, \citenamefont {Raimond}, \citenamefont
  {Haroche}, \citenamefont {Lutterbach},\ and\ \citenamefont
  {Davidovich}}]{Nogues2000}%
  \BibitemOpen
  \bibfield  {author} {\bibinfo {author} {\bibfnamefont {G.}~\bibnamefont
  {Nogues}}, \bibinfo {author} {\bibfnamefont {A.}~\bibnamefont
  {Rauschenbeutel}}, \bibinfo {author} {\bibfnamefont {S.}~\bibnamefont
  {Osnaghi}}, \bibinfo {author} {\bibfnamefont {P.}~\bibnamefont {Bertet}},
  \bibinfo {author} {\bibfnamefont {M.}~\bibnamefont {Brune}}, \bibinfo
  {author} {\bibfnamefont {J.~M.}\ \bibnamefont {Raimond}}, \bibinfo {author}
  {\bibfnamefont {S.}~\bibnamefont {Haroche}}, \bibinfo {author} {\bibfnamefont
  {L.~G.}\ \bibnamefont {Lutterbach}},\ and\ \bibinfo {author} {\bibfnamefont
  {L.}~\bibnamefont {Davidovich}},\ }\bibfield  {title} {\bibinfo {title}
  {{Measurement of a negative value for the Wigner function of radiation}},\
  }\href {https://doi.org/10.1103/PhysRevA.62.054101} {\bibfield  {journal}
  {\bibinfo  {journal} {Phys. Rev. A}\ }\textbf {\bibinfo {volume} {62}},\
  \bibinfo {pages} {054101} (\bibinfo {year} {2000})}\BibitemShut {NoStop}%
\bibitem [{\citenamefont {Bertet}\ \emph {et~al.}(2002)\citenamefont {Bertet},
  \citenamefont {Auffeves}, \citenamefont {Maioli}, \citenamefont {Osnaghi},
  \citenamefont {Meunier}, \citenamefont {Brune}, \citenamefont {Raimond},\
  and\ \citenamefont {Haroche}}]{Bertet2002}%
  \BibitemOpen
  \bibfield  {author} {\bibinfo {author} {\bibfnamefont {P.}~\bibnamefont
  {Bertet}}, \bibinfo {author} {\bibfnamefont {A.}~\bibnamefont {Auffeves}},
  \bibinfo {author} {\bibfnamefont {P.}~\bibnamefont {Maioli}}, \bibinfo
  {author} {\bibfnamefont {S.}~\bibnamefont {Osnaghi}}, \bibinfo {author}
  {\bibfnamefont {T.}~\bibnamefont {Meunier}}, \bibinfo {author} {\bibfnamefont
  {M.}~\bibnamefont {Brune}}, \bibinfo {author} {\bibfnamefont {J.~M.}\
  \bibnamefont {Raimond}},\ and\ \bibinfo {author} {\bibfnamefont
  {S.}~\bibnamefont {Haroche}},\ }\bibfield  {title} {\bibinfo {title} {{Direct
  Measurement of the Wigner Function of a One-Photon Fock State in a Cavity}},\
  }\href {https://doi.org/10.1103/PhysRevLett.89.200402} {\bibfield  {journal}
  {\bibinfo  {journal} {Phys. Rev. Lett.}\ }\textbf {\bibinfo {volume} {89}},\
  \bibinfo {pages} {200402} (\bibinfo {year} {2002})}\BibitemShut {NoStop}%
\bibitem [{\citenamefont {Vlastakis}\ \emph {et~al.}(2013)\citenamefont
  {Vlastakis}, \citenamefont {Kirchmair}, \citenamefont {Leghtas},
  \citenamefont {Nigg}, \citenamefont {Frunzio}, \citenamefont {Girvin},
  \citenamefont {Mirrahimi}, \citenamefont {Devoret},\ and\ \citenamefont
  {Schoelkopf}}]{Vlastakis:2013}%
  \BibitemOpen
  \bibfield  {author} {\bibinfo {author} {\bibfnamefont {B.}~\bibnamefont
  {Vlastakis}}, \bibinfo {author} {\bibfnamefont {G.}~\bibnamefont
  {Kirchmair}}, \bibinfo {author} {\bibfnamefont {Z.}~\bibnamefont {Leghtas}},
  \bibinfo {author} {\bibfnamefont {S.~E.}\ \bibnamefont {Nigg}}, \bibinfo
  {author} {\bibfnamefont {L.}~\bibnamefont {Frunzio}}, \bibinfo {author}
  {\bibfnamefont {S.~M.}\ \bibnamefont {Girvin}}, \bibinfo {author}
  {\bibfnamefont {M.}~\bibnamefont {Mirrahimi}}, \bibinfo {author}
  {\bibfnamefont {M.~H.}\ \bibnamefont {Devoret}},\ and\ \bibinfo {author}
  {\bibfnamefont {R.~J.}\ \bibnamefont {Schoelkopf}},\ }\bibfield  {title}
  {\bibinfo {title} {{Deterministically Encoding Quantum Information Using
  100-Photon Schr{\"o}dinger Cat States}},\ }\href
  {https://doi.org/10.1126/science.1243289} {\bibfield  {journal} {\bibinfo
  {journal} {Science}\ }\textbf {\bibinfo {volume} {342}},\ \bibinfo {pages}
  {607} (\bibinfo {year} {2013})}\BibitemShut {NoStop}%
\bibitem [{\citenamefont {Kurtsiefer}\ \emph {et~al.}(1997)\citenamefont
  {Kurtsiefer}, \citenamefont {Pfau},\ and\ \citenamefont
  {Mlynek}}]{Kurtsiefer1997}%
  \BibitemOpen
  \bibfield  {author} {\bibinfo {author} {\bibfnamefont {C.}~\bibnamefont
  {Kurtsiefer}}, \bibinfo {author} {\bibfnamefont {T.}~\bibnamefont {Pfau}},\
  and\ \bibinfo {author} {\bibfnamefont {J.}~\bibnamefont {Mlynek}},\
  }\bibfield  {title} {\bibinfo {title} {{Measurement of the Wigner function of
  an ensemble of helium atoms}},\ }\href {https://doi.org/10.1038/386150a0}
  {\bibfield  {journal} {\bibinfo  {journal} {Nature}\ }\textbf {\bibinfo
  {volume} {386}},\ \bibinfo {pages} {150} (\bibinfo {year}
  {1997})}\BibitemShut {NoStop}%
\bibitem [{\citenamefont {Brown}\ \emph {et~al.}(2022)\citenamefont {Brown},
  \citenamefont {Muleady}, \citenamefont {Dworschack}, \citenamefont
  {Lewis-Swan}, \citenamefont {Rey}, \citenamefont {Romero-Isart},\ and\
  \citenamefont {Regal}}]{Brown2022}%
  \BibitemOpen
  \bibfield  {author} {\bibinfo {author} {\bibfnamefont {M.~O.}\ \bibnamefont
  {Brown}}, \bibinfo {author} {\bibfnamefont {S.~R.}\ \bibnamefont {Muleady}},
  \bibinfo {author} {\bibfnamefont {W.~J.}\ \bibnamefont {Dworschack}},
  \bibinfo {author} {\bibfnamefont {R.~J.}\ \bibnamefont {Lewis-Swan}},
  \bibinfo {author} {\bibfnamefont {A.~M.}\ \bibnamefont {Rey}}, \bibinfo
  {author} {\bibfnamefont {O.}~\bibnamefont {Romero-Isart}},\ and\ \bibinfo
  {author} {\bibfnamefont {C.~A.}\ \bibnamefont {Regal}},\ }\href
  {https://doi.org/10.48550/arxiv.2203.03053} {\bibinfo {title}
  {{Time-of-Flight Quantum Tomography of Single Atom Motion}}} (\bibinfo {year}
  {2022})\BibitemShut {NoStop}%
\bibitem [{\citenamefont {Deutsch}\ and\ \citenamefont
  {Jessen}(1998)}]{Deutsch:1998}%
  \BibitemOpen
  \bibfield  {author} {\bibinfo {author} {\bibfnamefont {I.~H.}\ \bibnamefont
  {Deutsch}}\ and\ \bibinfo {author} {\bibfnamefont {P.~S.}\ \bibnamefont
  {Jessen}},\ }\bibfield  {title} {\bibinfo {title} {{Quantum-state control in
  optical lattices}},\ }\href {https://doi.org/10.1103/PhysRevA.57.1972}
  {\bibfield  {journal} {\bibinfo  {journal} {Phys. Rev. A}\ }\textbf {\bibinfo
  {volume} {57}},\ \bibinfo {pages} {1972} (\bibinfo {year}
  {1998})}\BibitemShut {NoStop}%
\bibitem [{\citenamefont {Robens}\ \emph {et~al.}(2018)\citenamefont {Robens},
  \citenamefont {Brakhane}, \citenamefont {Alt}, \citenamefont {Meschede},
  \citenamefont {Zopes},\ and\ \citenamefont {Alberti}}]{Robens2018}%
  \BibitemOpen
  \bibfield  {author} {\bibinfo {author} {\bibfnamefont {C.}~\bibnamefont
  {Robens}}, \bibinfo {author} {\bibfnamefont {S.}~\bibnamefont {Brakhane}},
  \bibinfo {author} {\bibfnamefont {W.}~\bibnamefont {Alt}}, \bibinfo {author}
  {\bibfnamefont {D.}~\bibnamefont {Meschede}}, \bibinfo {author}
  {\bibfnamefont {J.}~\bibnamefont {Zopes}},\ and\ \bibinfo {author}
  {\bibfnamefont {A.}~\bibnamefont {Alberti}},\ }\bibfield  {title} {\bibinfo
  {title} {{Fast, High-Precision Optical Polarization Synthesizer for
  Ultracold-Atom Experiments}},\ }\href
  {https://doi.org/10.1103/PhysRevApplied.9.034016} {\bibfield  {journal}
  {\bibinfo  {journal} {Phys. Rev. Appl.}\ }\textbf {\bibinfo {volume} {9}},\
  \bibinfo {pages} {034016} (\bibinfo {year} {2018})}\BibitemShut {NoStop}%
\bibitem [{\citenamefont {Weidner}\ and\ \citenamefont
  {Anderson}(2018)}]{Weidner:2018}%
  \BibitemOpen
  \bibfield  {author} {\bibinfo {author} {\bibfnamefont {C.~A.}\ \bibnamefont
  {Weidner}}\ and\ \bibinfo {author} {\bibfnamefont {D.~Z.}\ \bibnamefont
  {Anderson}},\ }\bibfield  {title} {\bibinfo {title} {{Experimental
  Demonstration of Shaken-Lattice Interferometry}},\ }\href
  {https://doi.org/10.1103/PhysRevLett.120.263201} {\bibfield  {journal}
  {\bibinfo  {journal} {Phys. Rev. Lett.}\ }\textbf {\bibinfo {volume} {120}},\
  \bibinfo {pages} {263201} (\bibinfo {year} {2018})}\BibitemShut {NoStop}%
\bibitem [{\citenamefont {Modugno}\ and\ \citenamefont
  {Pettini}(2012)}]{Modugno:2012}%
  \BibitemOpen
  \bibfield  {author} {\bibinfo {author} {\bibfnamefont {M.}~\bibnamefont
  {Modugno}}\ and\ \bibinfo {author} {\bibfnamefont {G.}~\bibnamefont
  {Pettini}},\ }\bibfield  {title} {\bibinfo {title} {{Maximally localized
  Wannier functions for ultracold atoms in one-dimensional double-well periodic
  potentials}},\ }\href {https://doi.org/10.1088/1367-2630/14/5/055004}
  {\bibfield  {journal} {\bibinfo  {journal} {New J. Phys.}\ }\textbf {\bibinfo
  {volume} {14}},\ \bibinfo {pages} {055004} (\bibinfo {year}
  {2012})}\BibitemShut {NoStop}%
\bibitem [{\citenamefont {Blatt}\ \emph {et~al.}(2009)\citenamefont {Blatt},
  \citenamefont {Thomsen}, \citenamefont {Campbell}, \citenamefont {Ludlow},
  \citenamefont {Swallows}, \citenamefont {Martin}, \citenamefont {Boyd},\ and\
  \citenamefont {Ye}}]{Blatt:2009}%
  \BibitemOpen
  \bibfield  {author} {\bibinfo {author} {\bibfnamefont {S.}~\bibnamefont
  {Blatt}}, \bibinfo {author} {\bibfnamefont {J.~W.}\ \bibnamefont {Thomsen}},
  \bibinfo {author} {\bibfnamefont {G.~K.}\ \bibnamefont {Campbell}}, \bibinfo
  {author} {\bibfnamefont {A.~D.}\ \bibnamefont {Ludlow}}, \bibinfo {author}
  {\bibfnamefont {M.~D.}\ \bibnamefont {Swallows}}, \bibinfo {author}
  {\bibfnamefont {M.~J.}\ \bibnamefont {Martin}}, \bibinfo {author}
  {\bibfnamefont {M.~M.}\ \bibnamefont {Boyd}},\ and\ \bibinfo {author}
  {\bibfnamefont {J.}~\bibnamefont {Ye}},\ }\bibfield  {title} {\bibinfo
  {title} {{Rabi spectroscopy and excitation inhomogeneity in a one-dimensional
  optical lattice clock}},\ }\href {https://doi.org/10.1103/PhysRevA.80.052703}
  {\bibfield  {journal} {\bibinfo  {journal} {Phys. Rev. A}\ }\textbf {\bibinfo
  {volume} {80}},\ \bibinfo {pages} {052703} (\bibinfo {year}
  {2009})}\BibitemShut {NoStop}%
\bibitem [{\citenamefont {Lam}\ \emph {et~al.}(2021)\citenamefont {Lam},
  \citenamefont {Peter}, \citenamefont {Groh}, \citenamefont {Alt},
  \citenamefont {Robens}, \citenamefont {Meschede}, \citenamefont {Negretti},
  \citenamefont {Montangero}, \citenamefont {Calarco},\ and\ \citenamefont
  {Alberti}}]{Lam2021a}%
  \BibitemOpen
  \bibfield  {author} {\bibinfo {author} {\bibfnamefont {M.~R.}\ \bibnamefont
  {Lam}}, \bibinfo {author} {\bibfnamefont {N.}~\bibnamefont {Peter}}, \bibinfo
  {author} {\bibfnamefont {T.}~\bibnamefont {Groh}}, \bibinfo {author}
  {\bibfnamefont {W.}~\bibnamefont {Alt}}, \bibinfo {author} {\bibfnamefont
  {C.}~\bibnamefont {Robens}}, \bibinfo {author} {\bibfnamefont
  {D.}~\bibnamefont {Meschede}}, \bibinfo {author} {\bibfnamefont
  {A.}~\bibnamefont {Negretti}}, \bibinfo {author} {\bibfnamefont
  {S.}~\bibnamefont {Montangero}}, \bibinfo {author} {\bibfnamefont
  {T.}~\bibnamefont {Calarco}},\ and\ \bibinfo {author} {\bibfnamefont
  {A.}~\bibnamefont {Alberti}},\ }\bibfield  {title} {\bibinfo {title}
  {{Demonstration of Quantum Brachistochrones between Distant States of an
  Atom}},\ }\href {https://doi.org/10.1103/PhysRevX.11.011035} {\bibfield
  {journal} {\bibinfo  {journal} {Phys. Rev. X}\ }\textbf {\bibinfo {volume}
  {11}},\ \bibinfo {pages} {011035} (\bibinfo {year} {2021})}\BibitemShut
  {NoStop}%
\bibitem [{\citenamefont {Schlosser}\ \emph {et~al.}(2001)\citenamefont
  {Schlosser}, \citenamefont {Reymond}, \citenamefont {Protsenko},\ and\
  \citenamefont {Grangier}}]{Schlosser:2001}%
  \BibitemOpen
  \bibfield  {author} {\bibinfo {author} {\bibfnamefont {N.}~\bibnamefont
  {Schlosser}}, \bibinfo {author} {\bibfnamefont {G.}~\bibnamefont {Reymond}},
  \bibinfo {author} {\bibfnamefont {I.}~\bibnamefont {Protsenko}},\ and\
  \bibinfo {author} {\bibfnamefont {P.}~\bibnamefont {Grangier}},\ }\bibfield
  {title} {\bibinfo {title} {{Sub-poissonian loading of single atoms in a
  microscopic dipole trap}},\ }\href {https://doi.org/10.1038/35082512}
  {\bibfield  {journal} {\bibinfo  {journal} {Nature}\ }\textbf {\bibinfo
  {volume} {411}},\ \bibinfo {pages} {1024} (\bibinfo {year}
  {2001})}\BibitemShut {NoStop}%
\bibitem [{\citenamefont {Belmechri}\ \emph {et~al.}(2013)\citenamefont
  {Belmechri}, \citenamefont {F{\"o}rster}, \citenamefont {Alt}, \citenamefont
  {Widera}, \citenamefont {Meschede},\ and\ \citenamefont
  {Alberti}}]{Belmechri2013}%
  \BibitemOpen
  \bibfield  {author} {\bibinfo {author} {\bibfnamefont {N.}~\bibnamefont
  {Belmechri}}, \bibinfo {author} {\bibfnamefont {L.}~\bibnamefont
  {F{\"o}rster}}, \bibinfo {author} {\bibfnamefont {W.}~\bibnamefont {Alt}},
  \bibinfo {author} {\bibfnamefont {A.}~\bibnamefont {Widera}}, \bibinfo
  {author} {\bibfnamefont {D.}~\bibnamefont {Meschede}},\ and\ \bibinfo
  {author} {\bibfnamefont {A.}~\bibnamefont {Alberti}},\ }\bibfield  {title}
  {\bibinfo {title} {{Microwave control of atomic motional states in a
  spin-dependent optical lattice}},\ }\href
  {https://doi.org/10.1088/0953-4075/46/10/104006} {\bibfield  {journal}
  {\bibinfo  {journal} {J. Phys. B: At. Mol. Phys.}\ }\textbf {\bibinfo
  {volume} {46}},\ \bibinfo {pages} {104006} (\bibinfo {year}
  {2013})}\BibitemShut {NoStop}%
\bibitem [{\citenamefont {Ness}\ \emph {et~al.}(2021)\citenamefont {Ness},
  \citenamefont {Lam}, \citenamefont {Alt}, \citenamefont {Meschede},
  \citenamefont {Sagi},\ and\ \citenamefont {Alberti}}]{Ness:2021}%
  \BibitemOpen
  \bibfield  {author} {\bibinfo {author} {\bibfnamefont {G.}~\bibnamefont
  {Ness}}, \bibinfo {author} {\bibfnamefont {M.~R.}\ \bibnamefont {Lam}},
  \bibinfo {author} {\bibfnamefont {W.}~\bibnamefont {Alt}}, \bibinfo {author}
  {\bibfnamefont {D.}~\bibnamefont {Meschede}}, \bibinfo {author}
  {\bibfnamefont {Y.}~\bibnamefont {Sagi}},\ and\ \bibinfo {author}
  {\bibfnamefont {A.}~\bibnamefont {Alberti}},\ }\bibfield  {title} {\bibinfo
  {title} {{Observing crossover between quantum speed limits}},\ }\href
  {https://doi.org/10.1126/sciadv.abj9119} {\bibfield  {journal} {\bibinfo
  {journal} {Sci. Adv.}\ }\textbf {\bibinfo {volume} {7}},\ \bibinfo {pages}
  {eabj9119} (\bibinfo {year} {2021})}\BibitemShut {NoStop}%
\bibitem [{\citenamefont {Kuhr}\ \emph {et~al.}(2003)\citenamefont {Kuhr},
  \citenamefont {Alt}, \citenamefont {Schrader}, \citenamefont {Dotsenko},
  \citenamefont {Miroshnychenko}, \citenamefont {Rosenfeld}, \citenamefont
  {Khudaverdyan}, \citenamefont {Gomer}, \citenamefont {Rauschenbeutel},\ and\
  \citenamefont {Meschede}}]{Kuhr:2003}%
  \BibitemOpen
  \bibfield  {author} {\bibinfo {author} {\bibfnamefont {S.}~\bibnamefont
  {Kuhr}}, \bibinfo {author} {\bibfnamefont {W.}~\bibnamefont {Alt}}, \bibinfo
  {author} {\bibfnamefont {D.}~\bibnamefont {Schrader}}, \bibinfo {author}
  {\bibfnamefont {I.}~\bibnamefont {Dotsenko}}, \bibinfo {author}
  {\bibfnamefont {Y.}~\bibnamefont {Miroshnychenko}}, \bibinfo {author}
  {\bibfnamefont {W.}~\bibnamefont {Rosenfeld}}, \bibinfo {author}
  {\bibfnamefont {M.}~\bibnamefont {Khudaverdyan}}, \bibinfo {author}
  {\bibfnamefont {V.}~\bibnamefont {Gomer}}, \bibinfo {author} {\bibfnamefont
  {A.}~\bibnamefont {Rauschenbeutel}},\ and\ \bibinfo {author} {\bibfnamefont
  {D.}~\bibnamefont {Meschede}},\ }\bibfield  {title} {\bibinfo {title}
  {{Coherence Properties and Quantum State Transportation in an Optical
  Conveyor Belt}},\ }\href {https://doi.org/10.1103/PhysRevLett.91.213002}
  {\bibfield  {journal} {\bibinfo  {journal} {Phys. Rev. Lett.}\ }\textbf
  {\bibinfo {volume} {91}},\ \bibinfo {pages} {213002} (\bibinfo {year}
  {2003})}\BibitemShut {NoStop}%
\bibitem [{\citenamefont {Ramola}\ \emph {et~al.}(2021)\citenamefont {Ramola},
  \citenamefont {Winkelmann}, \citenamefont {Chandrashekara}, \citenamefont
  {Alt}, \citenamefont {Xu}, \citenamefont {Meschede},\ and\ \citenamefont
  {Alberti}}]{Ramola2021a}%
  \BibitemOpen
  \bibfield  {author} {\bibinfo {author} {\bibfnamefont {G.}~\bibnamefont
  {Ramola}}, \bibinfo {author} {\bibfnamefont {R.}~\bibnamefont {Winkelmann}},
  \bibinfo {author} {\bibfnamefont {K.}~\bibnamefont {Chandrashekara}},
  \bibinfo {author} {\bibfnamefont {W.}~\bibnamefont {Alt}}, \bibinfo {author}
  {\bibfnamefont {P.}~\bibnamefont {Xu}}, \bibinfo {author} {\bibfnamefont
  {D.}~\bibnamefont {Meschede}},\ and\ \bibinfo {author} {\bibfnamefont
  {A.}~\bibnamefont {Alberti}},\ }\bibfield  {title} {\bibinfo {title} {{Ramsey
  Imaging of Optical Traps}},\ }\href
  {https://doi.org/10.1103/PhysRevApplied.16.024041} {\bibfield  {journal}
  {\bibinfo  {journal} {Phys. Rev. Appl.}\ }\textbf {\bibinfo {volume} {16}},\
  \bibinfo {pages} {024041} (\bibinfo {year} {2021})}\BibitemShut {NoStop}%
\bibitem [{\citenamefont {Robens}\ \emph
  {et~al.}(2017{\natexlab{a}})\citenamefont {Robens}, \citenamefont {Alt},
  \citenamefont {Emary}, \citenamefont {Meschede},\ and\ \citenamefont
  {Alberti}}]{Robens:2017}%
  \BibitemOpen
  \bibfield  {author} {\bibinfo {author} {\bibfnamefont {C.}~\bibnamefont
  {Robens}}, \bibinfo {author} {\bibfnamefont {W.}~\bibnamefont {Alt}},
  \bibinfo {author} {\bibfnamefont {C.}~\bibnamefont {Emary}}, \bibinfo
  {author} {\bibfnamefont {D.}~\bibnamefont {Meschede}},\ and\ \bibinfo
  {author} {\bibfnamefont {A.}~\bibnamefont {Alberti}},\ }\bibfield  {title}
  {\bibinfo {title} {{Atomic "bomb testing": the Elitzur-Vaidman experiment
  violates the Leggett-Garg inequality}},\ }\href
  {https://doi.org/10.1007/s00340-016-6581-y} {\bibfield  {journal} {\bibinfo
  {journal} {Appl. Phys. B: Lasers. Opt.}\ }\textbf {\bibinfo {volume} {123}},\
  \bibinfo {pages} {12} (\bibinfo {year} {2017}{\natexlab{a}})}\BibitemShut
  {NoStop}%
\bibitem [{\citenamefont {Wu}\ \emph {et~al.}(2019)\citenamefont {Wu},
  \citenamefont {Kumar}, \citenamefont {Giraldo},\ and\ \citenamefont
  {Weiss}}]{Wu:2019a}%
  \BibitemOpen
  \bibfield  {author} {\bibinfo {author} {\bibfnamefont {T.-Y.}\ \bibnamefont
  {Wu}}, \bibinfo {author} {\bibfnamefont {A.}~\bibnamefont {Kumar}}, \bibinfo
  {author} {\bibfnamefont {F.}~\bibnamefont {Giraldo}},\ and\ \bibinfo {author}
  {\bibfnamefont {D.~S.}\ \bibnamefont {Weiss}},\ }\bibfield  {title} {\bibinfo
  {title} {{Stern-Gerlach detection of neutral-atom qubits in a state-dependent
  optical lattice}},\ }\href {https://doi.org/10.1038/s41567-019-0478-8}
  {\bibfield  {journal} {\bibinfo  {journal} {Nat. Phys.}\ }\textbf {\bibinfo
  {volume} {15}},\ \bibinfo {pages} {538} (\bibinfo {year} {2019})}\BibitemShut
  {NoStop}%
\bibitem [{\citenamefont {Robens}\ \emph
  {et~al.}(2017{\natexlab{b}})\citenamefont {Robens}, \citenamefont {Brakhane},
  \citenamefont {Alt}, \citenamefont {Klei{\ss}ler}, \citenamefont {Meschede},
  \citenamefont {Moon}, \citenamefont {Ramola},\ and\ \citenamefont
  {Alberti}}]{Robens:2017a}%
  \BibitemOpen
  \bibfield  {author} {\bibinfo {author} {\bibfnamefont {C.}~\bibnamefont
  {Robens}}, \bibinfo {author} {\bibfnamefont {S.}~\bibnamefont {Brakhane}},
  \bibinfo {author} {\bibfnamefont {W.}~\bibnamefont {Alt}}, \bibinfo {author}
  {\bibfnamefont {F.}~\bibnamefont {Klei{\ss}ler}}, \bibinfo {author}
  {\bibfnamefont {D.}~\bibnamefont {Meschede}}, \bibinfo {author}
  {\bibfnamefont {G.}~\bibnamefont {Moon}}, \bibinfo {author} {\bibfnamefont
  {G.}~\bibnamefont {Ramola}},\ and\ \bibinfo {author} {\bibfnamefont
  {A.}~\bibnamefont {Alberti}},\ }\bibfield  {title} {\bibinfo {title} {{High
  numerical aperture (NA = 0.92) objective lens for imaging and addressing of
  cold atoms}},\ }\href {https://doi.org/10.1364/OL.42.001043} {\bibfield
  {journal} {\bibinfo  {journal} {Opt. Lett.}\ }\textbf {\bibinfo {volume}
  {42}},\ \bibinfo {pages} {1043} (\bibinfo {year}
  {2017}{\natexlab{b}})}\BibitemShut {NoStop}%
\bibitem [{\citenamefont {MacNamara}\ and\ \citenamefont
  {Strang}(2016)}]{MacNamara2016}%
  \BibitemOpen
  \bibfield  {author} {\bibinfo {author} {\bibfnamefont {S.}~\bibnamefont
  {MacNamara}}\ and\ \bibinfo {author} {\bibfnamefont {G.}~\bibnamefont
  {Strang}},\ }\bibfield  {title} {\bibinfo {title} {{Operator Splitting}},\
  }in\ \href {https://doi.org/10.1007/978-3-319-41589-5_3} {\emph {\bibinfo
  {booktitle} {{Splitting Methods in Communication, Imaging, Science, and
  Engineering}}}}\ (\bibinfo  {publisher} {{Springer International
  Publishing}},\ \bibinfo {year} {2016})\ pp.\ \bibinfo {pages}
  {95--114}\BibitemShut {NoStop}%
\bibitem [{\citenamefont {Barredo}\ \emph {et~al.}(2016)\citenamefont
  {Barredo}, \citenamefont {de~L{\'e}s{\'e}leuc}, \citenamefont {Lienhard},
  \citenamefont {Lahaye},\ and\ \citenamefont {Browaeys}}]{Barredo}%
  \BibitemOpen
  \bibfield  {author} {\bibinfo {author} {\bibfnamefont {D.}~\bibnamefont
  {Barredo}}, \bibinfo {author} {\bibfnamefont {S.}~\bibnamefont
  {de~L{\'e}s{\'e}leuc}}, \bibinfo {author} {\bibfnamefont {V.}~\bibnamefont
  {Lienhard}}, \bibinfo {author} {\bibfnamefont {T.}~\bibnamefont {Lahaye}},\
  and\ \bibinfo {author} {\bibfnamefont {A.}~\bibnamefont {Browaeys}},\
  }\bibfield  {title} {\bibinfo {title} {{An atom-by-atom assembler of
  defect-free arbitrary two-dimensional atomic arrays}},\ }\href
  {https://doi.org/10.1126/science.aah3778} {\bibfield  {journal} {\bibinfo
  {journal} {Science}\ }\textbf {\bibinfo {volume} {354}},\ \bibinfo {pages}
  {1021} (\bibinfo {year} {2016})}\BibitemShut {NoStop}%
\bibitem [{\citenamefont {Roberts}\ \emph {et~al.}(2014)\citenamefont
  {Roberts}, \citenamefont {McKellar}, \citenamefont {Fekete}, \citenamefont
  {Rakonjac}, \citenamefont {Deb},\ and\ \citenamefont
  {Kj{\ae}rgaard}}]{Kjaergaard}%
  \BibitemOpen
  \bibfield  {author} {\bibinfo {author} {\bibfnamefont {K.~O.}\ \bibnamefont
  {Roberts}}, \bibinfo {author} {\bibfnamefont {T.}~\bibnamefont {McKellar}},
  \bibinfo {author} {\bibfnamefont {J.}~\bibnamefont {Fekete}}, \bibinfo
  {author} {\bibfnamefont {A.}~\bibnamefont {Rakonjac}}, \bibinfo {author}
  {\bibfnamefont {A.~B.}\ \bibnamefont {Deb}},\ and\ \bibinfo {author}
  {\bibfnamefont {N.}~\bibnamefont {Kj{\ae}rgaard}},\ }\bibfield  {title}
  {\bibinfo {title} {{Steerable optical tweezers for ultracold atom studies}},\
  }\href {https://doi.org/10.1364/OL.39.002012} {\bibfield  {journal} {\bibinfo
   {journal} {Opt. Lett.}\ }\textbf {\bibinfo {volume} {39}},\ \bibinfo {pages}
  {2012} (\bibinfo {year} {2014})}\BibitemShut {NoStop}%
\bibitem [{\citenamefont {Henderson}\ \emph {et~al.}(2009)\citenamefont
  {Henderson}, \citenamefont {Ryu}, \citenamefont {MacCormick},\ and\
  \citenamefont {Boshier}}]{Henderson_2009}%
  \BibitemOpen
  \bibfield  {author} {\bibinfo {author} {\bibfnamefont {K.}~\bibnamefont
  {Henderson}}, \bibinfo {author} {\bibfnamefont {C.}~\bibnamefont {Ryu}},
  \bibinfo {author} {\bibfnamefont {C.}~\bibnamefont {MacCormick}},\ and\
  \bibinfo {author} {\bibfnamefont {M.~G.}\ \bibnamefont {Boshier}},\
  }\bibfield  {title} {\bibinfo {title} {{Experimental demonstration of
  painting arbitrary and dynamic potentials for Bose-Einstein condensates}},\
  }\href {https://doi.org/10.1088/1367-2630/11/4/043030} {\bibfield  {journal}
  {\bibinfo  {journal} {New J. Phys.}\ }\textbf {\bibinfo {volume} {11}},\
  \bibinfo {pages} {043030} (\bibinfo {year} {2009})}\BibitemShut {NoStop}%
\end{thebibliography}
\end{document}